\documentclass[a4paper]{article}
\usepackage{amsfonts}
\usepackage{amsmath}
\usepackage{graphicx}
\usepackage{graphics}
\usepackage{geometry}
\usepackage{float}
\usepackage{color}
\usepackage[hyperindex=true, colorlinks=true] {hyperref}
\usepackage{mathptmx}
\usepackage{newcent}
\usepackage{mathpazo}
\usepackage{setspace}
\usepackage{enumerate}
\usepackage{cases}
\newcommand{\edem}{\hfill $\Box$ }

\newcommand{\dsum}{\displaystyle\sum}

\newtheorem{prop}{Proposition}[section] 
 
\newtheorem{thm}{Theorem}[section] 

\newtheorem{definition}{Definition}[section] 
 
\newtheorem{rmqs}{Remarks}[section]

\newtheorem{Proof}{Proof}[section]

\title{Optimal Control of a Malaria Model with Long-Lasting Insecticide-Treated Nets}
\author{S. Y. Tchoumi $^{1}$\thanks{Corresponding author S. Y. Tchoumi email: sytchoumi83@gmail.com},  Y. T. Kouakep$^{2}$ ,  D. J. Fotsa Mbogne $^{1}$, J. C. Kamgang $^{1\dag}$ \thanks{Projet MASAIE INRIA Grand Est, France}\footnote{LIRIMA -- GRIMCAPE, Cameroun}, J. M. Tchuenche $^{3}$\\
\\
$^1$ Department of Mathematics and Computer Sciences\\ ENSAI -- University of N'Gaound\'er\'e, P. O. Box 455 N'Gaound\'er\'e (Cameroon)
\\ $^2$ Department of SFTI\\ EGCIM -- University of N'Gaound\'er\'e, P. O. Box 454 N'Gaound\'er\'e (Cameroon)
\\$^3$ School of Computer Science and Applied Mathematics\\ University of the
Witwatersrand, Johannesburg, South Africa}

\date{}

\begin{document}

\maketitle
    
\begin{abstract}
A deterministic multi-stage malaria model with a non-therapeutic control measure, the use of mosquito bednet is formulated and analyzed. The model basic reproduction number is derived, and analytical results show that the model’s equilibria are locally and globally asymptotically stable when certain threshold conditions are satisfied. Pontryagin's Maximum Principle with respect to a time dependent constant is used to derive the necessary conditions for the optimal usage of the Long-Lasting Insecticide-treated bed Nets (LLINs) to mitigate the malaria transmission dynamics. This is accomplished by introducing biologically admissible controls and $\epsilon\%$-approximate sub-optimal controls. The results from this study could help public health planners and policy decision-makers to design reachable and more practical malaria prevention programs "close" to the optimal strategy.
\end{abstract}



\section{Introduction}\label{sec.Intro}
 
Malaria is a vector-borne disease with high level of morbidity and mortality in the tropical regions of the globe. It is global public health concern. The disease is caused by several species of parasites of the Plasmodium genus type and transmitted to humans by the bite of a female anopheles mosquito when taking the blood meal necessary for egg production. In 2018, there was an estimated 228 million cases of malaria worldwide and the estimated number of deaths attributable to malaria amounted to 405,000 \cite{who}.
Various mathematical models of the transmission dynamics of malaria and its control have been proposed \cite{RAnguelov,chitso,DCollCZim,10.10382,syt1,Robrt03,jmt}. The very first model is that of Ross-MacDonald who laid the foundations for modeling malaria\cite{Macdo,Ross1911}. Models that include therapeutic (treatment and vaccination) and non-therapeutic (insecticide-treated bed net) measures have flourished in the literature \cite{chit_08, jmbo, jckt,ress}. Because insecticide-treated nets (ITNs) reduce human/mosquito contacts, distribution campaigns have been organized in affected countries, including Cameroon. However, the use of these mosquitoes treated bednets have not always been satisfactory as several people let holes in the bednets, do not use them every night or use these bednets for other activities such as fishing \cite{Short2018}. Mosquitoes insecticide-treated bednets could influence  the force of infection, the rate of recruitment of new females mosquito or the death rate of mosquitoes \cite{Agusto, chitso,buonomo, jmbo,jckt}. Moreoverm the use of these treated bednets can influence the rate of loss of immunity.
   
We formulate a mathematical model for the transmission dynamics of malaria in human populations, which takes the (good or bad) use of bednets as a control measure. First,we formulate the autonomous model with a constant proportion of bednets usage as control strategy. Next, we compute the basic reproduction number $\mathcal{R}_0$ and investigate the existence and stability of the equilibria. Analytical results show that both model equilibria; the disease-free and the endemic states are locally asymptotically stable when $\mathcal{R}_0<$ and when $\mathcal{R}_0>1$, respectively. However, the model could exhibit the phenomenon of backward bifurcation when $\mathcal{R}_0<1$, an epidemiological situation where, although necessary, having the basic reproduction number less than unity is not sufficient for malaria elimination \cite{jmt}.  
  
We then extend our autonomous model by considering a time-dependent control of the proportion of bednets usage. Optimal control theory is used to establish conditions under which the spread of malaria can mitigated. The characterization of the optimal control is obtained by the application of Pontryagin's maximum principle. We use numerical simulations to determine an optimal control strategy. In addition, we focus on a bednet control strategy since the other controls measures are expensive.  By other vector controls, we mean outdoor application of larvicides (chemical or biological), breeding habitat reduction (e.g., draining standing water), outdoor vector control (mosquito fogging, attractive toxic sugar bait (ATSB)), indoor residual spraying (IRS), repellents, including topical repellents, mosquito coils, etc, rapid diagnosis and treatment (RDT), preventative drugs like seasonal malaria chemo-prevention (SMC), intermittent preventative treatment (IPT) \cite{jckt}. Generally, the bednet control in the literature concerns the bednets usage, including insecticide-treated bed nets (ITNs), long-lasting insecticide-treated nets (LLINs), and untreated bednets (UBNs) \cite{jckt}.
   
The rest of the paper is organized as follows. In Section \ref{sec2}, we present the mathematical model for malaria transmission dynamics with a parameter $w$ that represents the proportion of persons having and using the treated mosquito bednets correctly. In Section \ref{sec3}, we propose an optimal control problem for the minimization of the number of infected humans while controlling the cost of control interventions with bednets. Finally, in Section \ref{sec4}, some numerical simulations are provided to support the analytical results and are interpreted from the epidemiological point of view. The last section is the conclusion.

\section{Model formulation and analysis }\label{sec2}
We denote with $b$ the proportion of people having a long-lasting insecticide-treated net and by $u$ the proportion of those who use it effectively. Therefore $w:=b \times u$ represents the proportion of people who own a mosquito net and use it adequately.

The number of bites on humans by a one female mosquito per day is $a$. the recruitment rate of mosquitoes  is $\Lambda_v$ and the rate of loss of immunity  is $\gamma_h$. The death rate of mosquitoes varies from a minimum value to a maximum value depending on whether the possession and use of the insecticide-treated mosquito net iare adequate or not. These different parameters are defined as follows: $a(w)=a_{max}-w\Delta_a$, $\Lambda_v(w)=\Lambda_v^{max}-w\Delta_{\Lambda_v}$, $\gamma_h(w)=\gamma_h^{min}+w\Delta_{\gamma_h}$ and $\mu_v(w)=\mu_v+w\Delta_{\mu_v}$, where $\Delta$ represents the difference between the maximum (max) value and the minimum (min) value of the indexed parameters above. 
\subsection{Model description and analysis }\label{moddesc}

We consider two populations namely human hosts and female mosquitoes that are homogeneously distributed. We also assume that female mosquitoes feed only on human blood. In the following sub-section, we will specifically describe the dynamics within the different populations.

\subsubsection{Host population structure and dynamics}

\begin{figure}[!h]
\begin{center}
\includegraphics[scale=0.5]{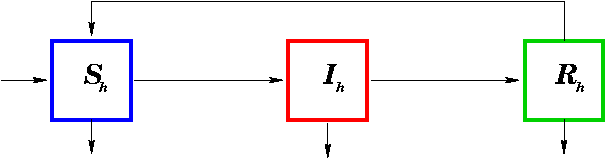} 
\caption{Dynamics of the human population}
\label{poph}
\end{center} 
\end{figure}

The human population is subdivided into three classes, namely the susceptible $ S_h $, the infectious $ I_h $ and the immune $ R_h $ as shown in Figure \ref{poph}. We have left the exposed compartment because its consideration or not will not influence the behavior of the evolution of the infection in the human population \cite{syt1}. When in contact with an infectious mosquito, a human can become infected at a rate of $ \alpha_{h} $ representing the force of infection. Infectious humans can gain immunity at a rate of $ \delta_h $, while the rate of loss of immunity is $ \gamma_h $. Recruitment is done only in the susceptible class at a rate of $ \Lambda_h $. In all compartments, there is an output of $ \mu_h $ due to natural death, in addition to which there is a disease-induced death rate in the infectious compartment.

\subsubsection{Mosquito Population Structure and Dynamics}

In the population of female mosquitoes, we consider two states of anopheles namely the active state (looking for the blood meal) and the resting state\cite{jmbo,jckt,syt2}. The new anopheles enter the compartment of susceptible in activity at rate of $\Lambda_v $. When contact with an infectious human, a susceptible mosquito can become infected at a rate of $ \alpha_{v} $ corresponding to the strength of infection of the mosquitoes. Once infected, the mosquito will go through seven phases of latency at rest and six questing activities before becoming infectious in activity and then infectious at rest.

\begin{figure}[!h]
\begin{center}
\includegraphics[scale=0.5]{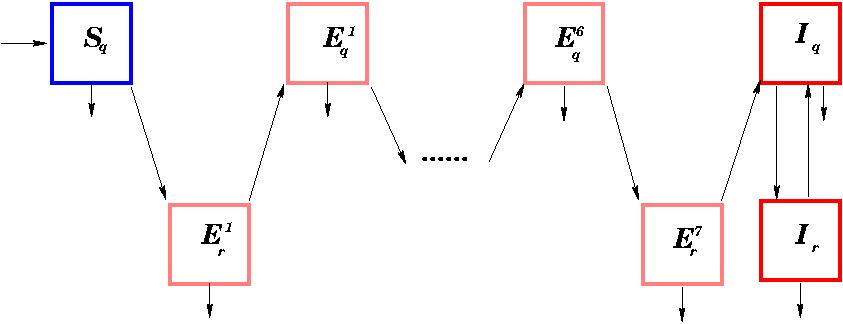} 
\caption{Dynamics of the mosquito population}
\label{schema1}
\end{center} 
\end{figure} 
 
\subsubsection{Model equation} 
The model variables and parameters values are presented in the Tables \ref{tab.var} and \ref{tab.fundpar}.

\begin{table}[htbp]
\begin{center}
\caption{Variable of model}\label{tab.var}
\begin{tabular}{ll}
\hline
Variable & Description \\
\hline 
humans & \\

$S_h$ & Number of susceptible humans within the  population \\
$I_h$ & Number of infectious humans within the  population\\
$R_h$ & Number of immune humans within the  population\\
\hline
mosquitoes & \\

$S_q$ & Number of questing susceptible mosquitoes \\
$E_q^{i}$ & Number of questing infected mosquitoes in step $i$\\
$E_r^{i}$ & Number of resting infected mosquitoes in step $i$\\
$I_q$ & Number of questing infectious mosquitoes  \\
$I_r$ & Number of resting infectious mosquitoes \\
\hline
\end{tabular}
\end{center}
\end{table}
 
 \begin{footnotesize}
\begin{center}
\begin{table}[htbp]
\caption{Fundamental model parameter}\label{tab.fundpar}
\begin{tabular}{llll}
\hline 
Parameter & Description & Value & Reference  \\ 
\hline 
human & & \\
$\Lambda_h$ & Immigration in the host population & $\frac{10000}{59*365}  $ & \cite{Agusto,buonomo}  \\ 
$\gamma_h^{max}$ &Max. transmission rate of loss of immunity within the host population & $ 0.0146$ &\cite{RAnguelov}\\  
$\gamma_h^{min}$ &Min. transmission rate of loss of immunity within the host population & $ 0.00055$ &\cite{RAnguelov} \\  
$\delta_h$ & Rate of recovery in the host population & $ 0.0035$ &  \cite{RAnguelov}\\ 
$\mu_h$ & Death rate in the host population & $\frac{1}{59 \times 365} $ & \cite{Agusto,buonomo}\\ 
$\mu_d$ & Disease-induced death rate within the host population & $ \left[10^{-5},10^{-3} \right] $ & \cite{Agusto}\\ 
$a_{max}$ & Max. number of bites on humans by one female mosquito per day & $19*0.5$ & \cite{RAnguelov}\\ 
$a_{min}$ &  Min. number of bites on humans by one female mosquito per day  & $4.3*0.33$ &\cite{RAnguelov}\\ 
$m$ & Infectivity coefficient of hosts due to a bite of infectious vector & $ 0.022$ & \cite{Chit_Hym_Cus_08}\\ 
\hline 
Mosquitoes & & \\
$\Lambda_v^{max} $ & Maximun immigration rate of vectors & $\frac{10^4}{21}+1$ & Assumed\\ 
$\Lambda_v^{min} $ & Minimun immigration rate of vectors & $\frac{10^4}{21}$ & \cite{buonomo}\\ 
$\chi $ & Rate at which resting vectors move to the questing state & $ \frac{1}{5}$ &  \cite{syt2}\\ 
$\beta $ & Rate at which questing vectors move to the resting state & $\frac{2}{3} $ &\cite{syt2} \\ 
$\mu_v=\mu_v^{min}$ & Natural death rate of vectors & $\frac{1}{21} $ & \cite{Agusto,buonomo}\\  
$\Delta_{\mu_v}$ & Death rate of vectors due to bednet & $\frac{1}{21}$ & \cite{Agusto,buonomo}\\ 
$c$ & Infectivity coefficient of vector due to a bite of  infectious host & $ 0.48 $  &\cite{Chit_Hym_Cus_08} \\ 
$\tilde{c} $ & Infectivity coefficient of vector due to a bite of  removed host group& $ 0.048 $  & \cite{Chit_Hym_Cus_08}\\ 
\hline
\end{tabular} 
\end{table}
\end{center}
\end{footnotesize}

\begin{table}[htbp]
\begin{center}
\caption{Derived model parameters}\label{tab.tabvd2}
\begin{tabular}{lll}
\hline
Parameter &  Formula &  Description  \\
\hline\\
$\alpha_h $ & $a(w)\dfrac{ m I_Q}{N_h}$  & Incidence rate of susceptible human  \\ \\
$\alpha_v$ & $a(w)\left( \dfrac{ c I_h}{N_h} +\dfrac{ \tilde{c} R_h}{N_h}\right) $ & Incidence rate of susceptible mosquitoes \\ \\
$f_r$ & $\dfrac{\chi}{\chi +\mu_v} $ & Resting frequency of mosquitoes  \\ \\
$f_q $ & $\dfrac{\beta}{\beta + \mu_v+w\Delta_{\mu_v} } $ & Questing frequency of mosquitoes \\ \\ 
\hline
\end{tabular}
\end{center}
\end{table}

Based on our model description and assumptions, we established the following system of non-linear ordinary differential equations \eqref{eqc1}.
\begin{equation}
\left\lbrace \begin{array}{l}
S_h^{'}=\Lambda_h+\gamma_h(w) R_h-(\alpha_{h}(w)+\mu_h)S_h, \\ \\
I_h^{'}= \alpha_{h} (w)S_h-(\delta_h+\mu_h+\mu_d) I_h,\\ \\
R_h^{'}=\delta_h I_h-(\gamma_h(w)+\mu_h)R_h ,\\ \\
S_q^{'}=\Lambda_v(w)-(\alpha_v(w)+\mu_v+w\Delta_{\mu_v})S_q, \\ \\
E^{{1}^{'}}_r=\alpha_v(w)S_q-(\chi+\mu_v)E^{1}_r, \\ \\
E^{{i}^{'}}_{q}=\chi E^{i}_{r}-(\beta+\mu_v+w\Delta_{\mu_v}) E^{i}_{q},\;\;\;\;\; 1 \leq i \leq 6, \\ \\
E^{{i}^{'}}_{r} =\beta E^{i-1}_{q}- (\chi+\mu_v) E^{i}_{r},\;\;\;\;\; 2 \leq i \leq 7,\\ \\
I_{r}^{'}=\beta I_q - (\chi+\mu_v)I_r,\\ \\
I_{q}^{'}=\chi (E_r^7+I_r)- (\beta+ \mu_v+w\Delta_{\mu_v})I_q.
\end{array} \right. \label{eqc1}
\end{equation}

\subsection{Well-Posedness, Dissipativity and Equilibria of the System}
System \eqref{eqc1} can be rewritten in matrix form as 
\begin{equation}
\mathbf{x}'=\mathbf{A}(\mathbf{x})\mathbf{x}+\mathbf{b} \Leftrightarrow \left\lbrace\begin{array}{lllllll}
\mathbf{x}_S' & = & \mathbf{A}_S(\mathbf{x})\mathbf{x}_S &  +& \mathbf{A}_{S,I}(\mathbf{x})\mathbf{x}_I & +& \mathbf{b}_S, \\ 
\mathbf{x}_I' & = & \mathbf{A}_I(\mathbf{x})\mathbf{x}_I, &  & & &
\end{array} \right.
\label{ecq2}
\end{equation}

where 

$$\mathbf{A}_{S}=\left( \begin{array}{cc}
-(\alpha_{h}+\mu_h) & 0 \\ 
0 & -(\alpha_v+\mu_v+w\Delta_{\mu_v})
\end{array}\right),  $$

$$\mathbf{A}_{SI}=\left(\begin{array}{ccccccccccccccccc}
0 & 0 & 0 & 0 & 0 & 0 & 0 & 0 & 0 & 0 & 0 & 0 & 0 & 0 & 0 & 0 & \gamma_h \\ 
0 & 0 & 0 & 0 & 0 & 0 & 0 & 0 & 0 & 0 & 0 & 0 & 0 & 0 & 0 & 0 & 0
\end{array} \right),$$
and
$$\mathbf{A}_{I}=\left(\begin{array}{cc}
\mathbf{A}_{11} & \mathbf{A}_{12} \\
\mathbf{A}_{21} & \mathbf{A}_{22}
\end{array} \right). $$

The $13 \times 13 $ matrix $\mathbf{A}_{11} $ is a 2-banded matrix whose diagonal and sub-diagonal elements are given by the vectors $d_0$ and $d_{-1}$ respectively, defined by
$$d_0=\left(  \underbrace{-(\chi + \mu_v),-(\beta+\mu_v+w\Delta_{\mu_v}),\ldots,-(\chi + \mu_v),-(\beta+\mu_v+w\Delta_{\mu_v})}_{12 components},-(\chi+\mu_v) \right),   $$ 
and 
$$d_{-1}=\left(  \underbrace{\chi,\beta,\ldots,\chi,\beta}_{12 components} \right).  $$
The matrix $\mathbf{A}_{12} $ is the $13 \times 4 $ matrix defined by
$$\mathbf{A}_{12}=\left(\begin{array}{cccc}
0 & 0 & \dfrac{acS_v^{*}}{N_h}&\dfrac{a\tilde{c}S_v^{*}}{N_h} \\ 
0 & 0 & 0 & 0\\ 
0 & 0 & 0 & 0 \\ 
0 & 0 & 0 & 0 \\ 
0 & 0 & 0 & 0\\ 
0 & 0 & 0 & 0\\ 
0 & 0 & 0 & 0\\ 
0 & 0 & 0 & 0\\ 
0 & 0 & 0 & 0\\ 
0 & 0 & 0 & 0\\ 
0 & 0 & 0 & 0\\ 
0 & 0 & 0& 0 \\
0 & 0 & 0& 0
\end{array}  \right). $$ 

The matrix $\mathbf{A}_{21} $ is the $4 \times 13 $ matrix defined by

$$\mathbf{A}_{21}=\left(\begin{array}{ccccccccccccc}
0 & 0 & 0 & 0 & 0 & 0 & 0 & 0 & 0 & 0 & 0 & 0 & 0 \\ 
0 & 0 & 0 & 0 & 0 & 0 & 0 & 0 & 0 & 0 & 0 & 0 & \chi \\ 
0 & 0 & 0 & 0 & 0 & 0 & 0 & 0 & 0 & 0 & 0 & 0 & 0\\
0 & 0 & 0 & 0 & 0 & 0 & 0 & 0 & 0 & 0 & 0 & 0 & 0
\end{array}  \right), $$ 
and the matrix $\mathbf{A}_{22} $ is the $4 \times 4 $ square matrix defined by
$$\mathbf{A}_{22}=\left(\begin{array}{cccc}
-(\chi + \mu_r) & \beta & 0 & 0 \\ 
\chi & -(\beta+\mu_v+w\Delta_{\mu_v}) & 0 & 0\\
0 & \dfrac{amS_v^{*}}{N_h^{*}} & -(\delta_h+\mu_h+\mu_d) & 0  \\ 
 0 & 0 &\delta_h  & -(\gamma_h+\mu_h)
\end{array}  \right). $$ 
\begin{prop}
The non-negative cone $R^{19}_+$ is positively invariant for system \eqref{ecq2}.
\end{prop}

\begin{Proof}
Let $a_{i j} (x)$ be the $(i, j)^{th}$ entry of $\mathbf{A(x)}$. Since $\mathbf{A(x)}$ is a Metzler matrix for all
$\mathbf{x} \in R^{19}_+$, it follows that in this region $a_{i j} (\mathbf{x}) \geq 0$ for all $i$ and $j$, $i \neq j$. The boundary of $ \mathbb{R}^{19}_+$ is the union of the sets $\mathcal{H}_i$, $i = 1 \ldots 19$, where $\mathcal{H}_i\equiv \left\lbrace  \mathbf{x} \in \mathbb{R}^{19}, \mid x_i=0\right\rbrace \cap \mathbb{R}^{19}_+ $.
Thus, for $x \in \mathcal{H}_i$,
$$x'_i=\sum_{j=1}^{19} a_{i j} (\mathbf{x})x_i +b_i=\sum_{j=1,i \neq j }^{19} a_{i j} (\mathbf{x})x_i +b_i \geq 0,$$
so that on the boundary of $R^{19}_+$, the tangents to all trajectories point within $\mathbb{R}^{19}_+$ . By continuity, it follows that all trajectories of system \eqref{ecq2} that begin inside $\mathbb{R}^{19}_+$ can never leave $\mathbb{R}^{19}_+$.

\end{Proof}

\begin{prop}
The simplex $\Omega=\left\lbrace \left(S_h,S_q,(E_r^i,E_q^i)_{1 \leq i \leq 6},E_r^7,I_r,I_q,I_h,R_h \right)\in \mathbb{R}^{19}_{+} /  0 \leq N_h \leq \dfrac{\Lambda_h}{\mu_h}, 0 \leq N_{v} \leq \dfrac{\Lambda_v(w)}{\mu_v}   \right\rbrace $ is a compact forward-invariant and absorbing set for system \eqref{eqc1}.
\end{prop}

\begin{Proof}
The following equations from system \eqref{eqc1} respectively describe the total population of humans and of vector.
\begin{equation}
\left\lbrace \begin{array}{l}
N_h^{'}=\Lambda_h-\mu_h N_h-\mu_d I_h, \\ \\
N_v^{'}=\Lambda_v(w)-\mu_v N_v-\tilde{\mu}_v w N_v^q.
\end{array} \right.
\end{equation}

We have $$\Lambda_h-(\mu_h +\mu_d)N_h \leq N_h^{'}\leq \Lambda_h-\mu_h N_h, $$ and $$\Lambda_v-(\mu_v+w\Delta_{\mu_v})N_v \leq N_v^{'}\leq \Lambda_v-\mu_v N_v.$$

Thus, $$\frac{\Lambda_h}{\mu_h +\mu_d}+\left(N_h(t_0)-\frac{\Lambda_h}{\mu_h+\mu_d}\right)e^{-(\mu_h+\mu_d) t} \leq N_h \leq \frac{\Lambda_h}{\mu_h}+\left(N_h(t_0)-\frac{\Lambda_h}{\mu_h}\right)e^{-\mu_h t},$$  and $$ \frac{\Lambda_v(w)}{\mu_v+w\Delta_{\mu_v}}+\left(N_v(t_0)-\frac{\Lambda_v(w)}{\mu_v+w\Delta_{\mu_v}}\right)e^{-(\mu_v+w\Delta_{\mu_v}) t} \leq N_v \leq \frac{\Lambda_v(w)}{\mu_v}+\left(N_v(t_0)-\frac{\Lambda_v(w)}{\mu_v}\right)e^{-\mu_v t}.$$

So, if $0 \leq N_h(t=0)\leq \frac{\Lambda_h}{\mu_h}$ and $0 \leq N_v(t=0)\leq \frac{\Lambda_v(w)}{\mu_v}$, then $\forall t \geq t_0$, $0 \leq N_h(t)\leq \frac{\Lambda_h}{\mu_h}$ and $0 \leq N_v(t)\leq \frac{\Lambda_v(w)}{\mu_v} \leq\frac{\Lambda_v^{max}}{\mu_v}$.
\end{Proof}

\subsubsection{Disease-free equilibrium}

\begin{thm}
System \eqref{eqc1} admits a disease-free equilibrium (DFE) given by $\mathbf{x}^{*}=(\mathbf{x}^{*}_S,\mathbf{x}^{*}_I)$ with $$\mathbf{x}^{*}_S=\left(\dfrac{\Lambda_h}{\mu_h},\dfrac{\Lambda_v(w)}{\mu_v+w\Delta_{\mu_v}}\right)$$ and $$\mathbf{x}^{*}_I=0_{\mathbb{R}^{17}} \in \mathbb{R}^{17}$$
\end{thm}

\begin{prop}
The system $\mathbf{x}' =  \mathbf{A}_S(\mathbf{x}^{*}).\left(\mathbf{x}-\mathbf{x}_S^{*}\right) $ is Globally Asymptotically Stable (GAS) at $\mathbf{x}^{*}_S$ on $\mathbb{R}^{2}_+ $.
\end{prop}

\begin{Proof}
The proof is immediate since $$\mathbf{A}_{S}(\mathbf{x}^{*})=\left( \begin{array}{cc}
-\mu_h & 0 \\ 
0 & -(\mu_v+w\Delta_{\mu_v}) \end{array}\right).$$ 
\end{Proof}
\subsubsection{Computation of threshold condition}
In this subsection, we determine a stability threshold condition using a technique well described and used in \cite{jmbo,syt2}. In our case, this threshold can be biologically interpreted as the basic reproduction number $\mathcal{R}_0$ \cite{jckt}.
\begin{thm}
The basic reproduction number $\mathcal{R}_0 $ of the system \eqref{eqc1} is
\begin{equation}
\mathcal{R}_0=\frac{S_{v}^{*} a m (f_rf_q)^{7} }{\beta {\left(1 -f_qf_r \right)} }\dfrac{a\left[c(\gamma_h+\mu_h)+\tilde{c}\delta_h \right]}{N_{h}^{*}{\left(\delta_{h} + \mu_{h} + \mu_d\right)(\gamma_h+\mu_h)}}.
\end{equation}
\end{thm}
\begin{Proof}
Since the model system \eqref{eqc1} can be reduced to the infection-free sub-variety of $(\mathbb{R}^{2}_{+})$, the system has a unique equilibrium $\mathbf{x}_S^{\ast}$ that is GAS. We seek for conditions under which the matrix $\mathbf{A}_{I}(\mathbf{x}^{\ast})$, that is the sub-matrix of the Jacobian matrix of the system \eqref{ecq2} reduced to the infected sub-variety at the DFE is stable. 

This matrix $\mathbf{A}_{I}(\mathbf{x}^{\ast})$ is a Metzler matrix, so we must seek for conditions for which the matrix $\mathbf{A}_{I}(\mathbf{x}{\ast}) $ is a Metzler stable matrix. We apply the algorithm given in \cite{KamSal07} to the matrix $\mathbf{A}_{I}(\mathbf{x}{\ast})$; we have: $\mathbf{A}_{I}(\mathbf{x}{\ast})$ is Metzler stable matrix if and only if $\mathbf{A}_{22} $ and $\mathbf{N} = \mathbf{A}_{11} -\mathbf{A}_{12} \times \mathbf{A}_{22}^{-1} \times \mathbf{A}_{21} $ are Metzler stable.

Since the matrix $ \mathbf{A}_{22} $ is Metzler stable, we are now interested in the matrix $ \mathbf{N} = \mathbf{A}_{11} -\mathbf{A}_{12} \times \mathbf{A}_{22}^{-1} \times \mathbf{A}_{21}, $ where

$\mathbf{N}=\left(\begin{array}{cc}
\mathbf{N}_{11} & \mathbf{N}_{12}\\
\mathbf{N}_{21} & \mathbf{N}_{22}
\end{array}\right). $

$$\mathbf{N}_{11}=\left(\begin{array}{rr}
-(\chi + \mu_v) & 0 \\
\chi & -(\beta + \mu_v+w\Delta_{\mu_v})
\end{array}\right), $$

$$\mathbf{N}_{12}=\left(\begin{array}{rrrrrrrrrrr}
0&0 & 0 & 0 & 0 & 0 & 0 & 0 & 0 & 0 & \frac{S_{h}^{*} S_{v}^{*} a^{2} \chi m f_q \left[c(\gamma_h+\mu_h)+\tilde{c}\delta_h \right]}{N_{h}^{2} \beta{\left(1 -f_qf_r \right)}{ {\left(\delta_{h} + \mu_{h} + \mu_d\right)(\gamma_h+\mu_h)}}} \\
0 & 0 & 0 & 0 & 0 & 0 & 0 & 0 & 0 & 0 & 0
\end{array}\right), $$

$$\mathbf{N}_{21}=\left(\begin{array}{rr}
0 & \beta \\
0 & 0 \\
0 & 0 \\
0 & 0 \\
0 & 0 \\
0 & 0 \\
0 & 0 \\
0 & 0 \\
0 & 0 \\
0 & 0 \\
0 & 0
\end{array}\right), $$

\begin{tiny}
$$\mathbf{N}_{22}=\left(\begin{array}{rrrrrrrrrrr}
-\chi - \mu_{v} & 0 & 0 & 0 & 0 & 0 & 0 & 0 & 0 & 0 & 0  \\
\chi & -(\beta +\mu_v+w\Delta_{\mu_v}) & 0 & 0 & 0 & 0 & 0 & 0 & 0 & 0 & 0 \\
0 & \beta & -\chi - \mu_{v} & 0 & 0 & 0 & 0 & 0 & 0 & 0 & 0 \\
0 & 0 & \chi & -(\beta +\mu_v+w\Delta_{\mu_v}) & 0 & 0 & 0 & 0 & 0 & 0 & 0\\
0 & 0 & 0 & \beta & -\chi - \mu_{v} & 0 & 0 & 0 & 0 & 0 & 0\\
0 & 0 & 0 & 0 & \chi & -(\beta +\mu_v+w\Delta_{\mu_v}) & 0 & 0 & 0 & 0 & 0\\
0 & 0 & 0 & 0 & 0 & \beta & -\chi - \mu_{v} & 0 & 0 & 0 & 0 \\
0 & 0 & 0 & 0 & 0 & 0 & \chi & -(\beta +\mu_v+w\Delta_{\mu_v}) & 0 & 0 & 0\\
0 & 0 & 0 & 0 & 0 & 0 & 0 & \beta & -\chi - \mu_{v} & 0 & 0\\
0 & 0 & 0 & 0 & 0 & 0 & 0 & 0 & \chi & -(\beta +\mu_v+w\Delta_{\mu_v}) & 0 \\
0 & 0 & 0 & 0 & 0 & 0 & 0 & 0 & 0 & \beta & -\chi - \mu_{v}\\
\end{array}\right). $$
\end{tiny}
Since $ \mathbf{N}_{22} $ is Metzler stable, we can do another iteration and focus on the matrix

 $\mathbf{L}=\mathbf{N}_{11}-\mathbf{N}_{12}\times \mathbf{N}_{22}^{-1} \times \mathbf{N}_{21}$. We then obtain

$$\mathbf{L}=\left(\begin{array}{rr}
-(\chi + \mu_v) & \frac{S_{h}^{*} S_{v}^{*} a^{2}  m (f_rf_q)^{6} \left[c(\gamma_h+\mu_h)+\tilde{c}\delta_h \right]}{N_{h}^{2} { \left(1 -f_qf_r \right)} {\left(\chi + \mu_v\right)} {\left(\delta_{h} + \mu_{h} + \mu_d\right)(\gamma_h+\mu_h)}} \\
\chi & -(\beta + \mu_v+w\Delta_{\mu_v})
\end{array}\right). $$ 

Because $\mathbf{L}_{22} $ is negative, then  Metzler stable, the matrix $\mathbf{A}_{I}$ is Metzler stable if and only if $\mathbf{L}_{11}-\mathbf{L}_{12}\times \mathbf{L}_{22}^{-1} \times \mathbf{L}_{21} \leq 0$, that is
$$-(\chi+\mu_v)+\frac{S_{h}^{*} S_{v}^{*} a^{2} m (f_rf_q)^{6} \left[c(\gamma_h+\mu_h)+\tilde{c}\delta_h \right]}{N_{h}^{2} {\beta \left(1 -f_qf_r \right)} {\left(\chi + \mu_v\right)} {\left(\delta_{h} + \mu_{h} + \mu_{d}\right)(\gamma_h+\mu_h)}} \times \dfrac{1}{\beta + (\mu_v+w\Delta_{\mu_v})} \times \chi  \leq 0. $$

After some algebraic manipulations, we obtain the following condition $\frac{S_{h}^{*} S_{v}^{*} a^{2} m (f_rf_q)^{7} \left[c(\gamma_h+\mu_h)+\tilde{c}\delta_h \right]}{N_{h}^{{*}^{2}}\beta {\left(1 -f_qf_r \right)} {\left(\delta_{h} + \mu_{h} + \mu_{d}\right)(\gamma_h+\mu_h)}} \leq 1 $. \edem 

\end{Proof}
\begin{thm}
Let $\zeta=\dfrac{\mu_h}{\mu_h+\mu_d}$, the DFE is GAS in $\Omega$ when $\mathcal{R}_0 < \zeta $.
\end{thm}
\begin{Proof}
Our proof relies on Theorem 4.3 in \cite{KamSal07}, which establishes global asymptotic stability (GAS) for epidemiological systems that can be expressed in matrix form \eqref{ecq2}. 
The demonstration is completely similar to that made in \cite{syt2}.
\end{Proof}

\subsubsection{Endemic equilibrium}
\begin{thm}
There exists $\mathcal{R}_{-},\mathcal{R}_c,\mathcal{R}_{+}\in\mathbb{R}$ such that the model system \eqref{eqc1} has:
\begin{itemize}
\item[(a)] a unique endemic equilibrium  if $ \mathcal{R}_0>1$, 
\item[(b)] a unique endemic equilibrium if $\mathcal{R}_0=1$ and $\mathcal{R}_c < 1 $, 
\item[(c)] two endemic equilibria if $\mathcal{R}_c < \mathcal{R}_0 < min(1,\mathcal{R}_{-}) $ or $max(\mathcal{R}_{c},\mathcal{R}_{+})<\mathcal{R}_{0}<1 $,
\item[(d)] No endemic equilibrium elsewhere.
\end{itemize}
\end{thm}
\begin{Proof}
An endemic equilibrium is any non-zero and positive solution of the following system:

\begin{numcases}{}
\Lambda_h+\gamma_h R_h-(\alpha_{h}+\mu_h)S_h=0,\label{eqe1} \\ 
\alpha_{h} S_h-(\delta_h+\mu_h+\mu_d) I_h=0,\label{eqe2} \\ 
\delta_h I_h-(\gamma_h+\mu_h)R_h=0 , \label{eqe3} \\ 
\Lambda_v(w)-(\alpha_v+(\mu_v+w\Delta_{\mu_v}))S_q=0, \label{eqe4}\\ 
\alpha_vS_q-(\chi+\mu_v)E^{1}_r=0,\label{eqe5} \\ 
\chi E^{i}_{r}-(\beta+(\mu_v+w\Delta_{\mu_v})) E^{i}_{q}=0,\;\;\;\;\; 1 \leq i \leq 6, \label{eqe6}\\ 
\beta E^{i-1}_{q}- (\chi+\mu_v) E^{i}_{r}=0,\;\;\;\;\; 2 \leq i \leq 6, \label{eqe7}\\ 
\beta (E_{q}^{6}+I_q) - (\chi+\mu_v)I_r=0, \label{eqe8}\\ 
\chi I_r- (\beta+ (\mu_v+w\Delta_{\mu_v}))I_q=0. \label{eqe9}
\end{numcases} 

The equations \eqref{eqe2} and \eqref{eqe3} allow us to write $S_h^{\star}$ and $R_h^{\star}$ in function of $I_h^{\star}$ as  follows: $S_h^{\star}=\dfrac{\delta_h+ \mu_h + \mu_d}{\alpha_h^{\star}}I_h^{\star}$ and $R_h^{\star}=\dfrac{\delta_h}{\gamma_h + \mu_h}I_h^{\star}. $

To simplify the expressions, let $D=\delta_h + \mu_h + \mu_d $ , $C=\dfrac{\delta_h}{\gamma_h + \mu_h} $ and $F=\gamma_h C $.

By subsequently replacing $S_{h}^{\star}$ and $R_{h}^{\star}$ by their values ​​in \eqref{eqe1}, we then obtain the expression of $I_{h}^{\star}$ with respect to $\alpha_{h}^{\star}$.

We also have $\alpha_v^{\star}=\dfrac{a(cI_h^{\star}+\tilde{c}R_h^{\star})}{N_h^{\star}}=\dfrac{a(C\tilde{c}+c)\alpha_h^{\star}}{(C+1)\alpha_h^{\star}+D}.$ Using the equations \eqref{eqe4},\eqref{eqe5},\eqref{eqe6},\eqref{eqe7},\eqref{eqe8} and \eqref{eqe9}, we have

 $S_q^{\star}=\dfrac{\Lambda_v(w)}{\alpha_v^{\star}+(\mu_v+w\Delta_{\mu_v})}$, $E_r^{1 \star}=\dfrac{\alpha_v^{\star}S_q^{\star}}{\chi + \mu_v} $, $E_q^{i \star}=\dfrac{\chi E_r^{i \star}}{\beta +(\mu_v+w\Delta_{\mu_v})}\;\;for\;\;1\leq i \leq 6 $, $E_r^{i \star}=\dfrac{\beta E_q^{(i-1) \star}}{\chi +\mu_v}\;\;for\;\;2\leq i \leq 7 $; $I_q^{\star}=\dfrac{\chi f_q E_r^{7 \star}}{\beta(1-f_q f_r)} $ et $I_r^{\star}=\dfrac{\beta I_q^{\star}}{\chi + \mu_v}. $
 
So all our unknowns are expressed in terms of $\alpha_{h}^{\star} $ and it only remains to determine the value ​​of $\alpha_h^{\star} $.
 
By definition, we have $\alpha_h^{\star}=\dfrac{amI_q^{\star}}{N_h^{\star}}$ and by replacing $I_q^{\star}$ and $N_h^{\star}$ by their values, ​​and after simplification and re-arrangement, we obtain
 
 \begin{equation}
 \alpha_h^{\star}\left[P_2 (\alpha_h^{\star})^2 + P_1 \alpha_h^{\star} + P_0 \right]=0 ,\label{Esd}
 \end{equation}
 
where
 $$P_2= -\beta^6 \chi^7 \mu_h\left(C+1 \right) \left( 1-f_q f_r \right) \left[a(\tilde{c}C+c)+(\mu_v+w\Delta_{\mu_v})(C+1) \right]<0,$$ 
 \begin{equation}
\begin{split}
P_1&=D\beta^6\chi^7\left(1-f_qf_r \right)\left[(\mu_v+w\Delta_{\mu_v})\mathcal{R}_0(D-F)-\mu_h\left(C(a\tilde{c}+(\mu_v+w\Delta_{\mu_v}))+ac+(\mu_v+w\Delta_{\mu_v}) \right) \right]\\
&=\dfrac{D\beta^6\chi^7\left(1-f_qf_r \right)}{(\mu_v+w\Delta_{\mu_v})(D-F)} \left[\mathcal{R}_0 - \mathcal{R}_c\right]\;\;with\;\;\mathcal{R}_c=\dfrac{\mu_h\left[C(a\tilde{c}+(\mu_v+w\Delta_{\mu_v}))+ac + (\mu_v+w\Delta_{\mu_v}) \right]}{(\mu_v+w\Delta_{\mu_v})(D-F)},
\end{split}
\end{equation} 
 $$P_0=D^2\mu_h (\mu_v+w\Delta_{\mu_v}) \beta^6 \chi^7 \left( \mathcal{R}_0-1 \right). $$ 
 
Equation \eqref{Esd} has solution $\alpha_h^{\star}=0$ and solutions of the equation $(E): P_2 (\alpha_h^{\star})^2 + P_1 \alpha_h^{\star} + P_0 = 0$.

The case $\alpha_h^{\star}=0 $ leads us to equilibrium without disease, we are interested in the equation $(E)$, of which we are going to analyze the number of positive solutions as a function of the value of $ \mathcal{R}_0 $.

\begin{enumerate}
\item If $\mathcal{R}_0>1 $ then, $P_0>0 $ and since $P_2<0 $, the discriminant $\Delta=P_1^2 -4 P_2 P_0 $ of the equation $(E)$ is positive, hence the equation $(E)$ has two real solutions. In addition, the product of the solutions is $p=\dfrac{P_0}{P_2}<0$. Hence, equation $(E)$ has a unique positive solution.
\item if $\mathcal{R}_0=1$, then, equation $(E)$ has two real solutions, which are zero and $-\dfrac{P_1}{P_2}$. But $P_2<0 $ so this solution is positive if $P_1>0 $, that is to say if $\mathcal{R}_0 > \mathcal{R}_c $.
\item if $\mathcal{R}_0<1 $ and $\Delta=P_{1}^2-4P_{2}P_{0}>0 $ and $\mathcal{R}_0 > \mathcal{R}_c $, then, equation $(E)$ admits two positive solutions.
\end{enumerate}

Let $P_2=-b_2$ $P_1=b_1(\mathcal{R}_0-\mathcal{R}_c) $ and $P_0=b_0(\mathcal{R}_0-1)$; $b_2,\;\;b_1\;\;and\;\;b_0$ are all positive coefficients. We have, 
$\Delta=P_1^2-4P_2P_0=b_1^2\mathcal{R}_0^{2}-(2b_1^{2}\mathcal{R}_c-4b_0b_2)\mathcal{R}_0 - 4b_0b_2+b_1\mathcal{R}_c^{2} $. 

The last condition can be re-written as follows $\left\lbrace \begin{array}{l}
\mathcal{R}_c < \mathcal{R}_0 <1,\\
\Delta=b_1^2\mathcal{R}_0^{2}-(2b_1^{2}\mathcal{R}_c-4b_0b_2)\mathcal{R}_0 - 4b_0b_2+b_1\mathcal{R}_c^{2}>0.
\end{array}\right. $

Let us study the sign of $\Delta $ in relation to the values of $\mathcal{R}_0$. Consider the equation $$(E_{\mathcal{R}_0}):b_1^2\mathcal{R}_0^{2}-(2b_1^{2}\mathcal{R}_c-4b_0b_2)\mathcal{R}_0 - 4b_0b_2+b_1^{2}\mathcal{R}_c^{2}=0$$

$(E_{\mathcal{R}_0})$ has as discriminant $\Delta_r=(2b_1^{2}\mathcal{R}_c-4b_0b_2)^2-4b_1^2(-4b_0b_2+b_1^{2}\mathcal{R}_c^{2} )=16b_2b_0\left[ b_2b_0+b_1^{2}(1-\mathcal{R}_c)\right] $ which is positive for $\mathcal{R}_c<1$, and the equation $(E_{\mathcal{R}_0})$ has two solutions $\mathcal{R}_{-} $ and $\mathcal{R}_{+}$. 

We then have $\left\lbrace \begin{array}{l}
\mathcal{R}_c < \mathcal{R}_0 <1,\\
\mathcal{R}_0 \in \left]-\infty, \mathcal{R}_{-} \right[ \cup \left]\mathcal{R}_{+},+\infty \right[,
\end{array}\right. $

which yields $\mathcal{R}_c < \mathcal{R}_0 < min(1,\mathcal{R}_{-}) $ where $max(\mathcal{R}_{c},\mathcal{R}_{+})<\mathcal{R}_{0}<1 $.
\end{Proof}

\begin{rmqs}
For the study of the (global) stability of the endemic equilibrium, one could follow the approach in \cite{jckt} by using a suitable Lyapunuv like functional along the positive flow of the model \ref{eqc1} on a "two domains" subdivision of the phase state $\mathbb{R}^{19}$, under appropriate conditions. 
\end{rmqs}

\subsubsection{Existence of backward bifurcation}\label{bf}

Note that the disease-free equilibrium is only globally asymptotically stable when $ \mathcal{R}_0 <\zeta <1 $, so it is possible that if this condition is violated, bistability could occur. That is, for $ \zeta < \mathcal{R}_0 <1 $, a stable DFE could co-exist with a stable endemic equilibrium, a phenomenon known as backward bifurcation \cite{AleMog06, MR2018867, DusHu98, hernandez, jmt}.

\section{Optimal control model}\label{sec3}
There are several methods to mitigate the prevalence of malaria in a community by reducing the biological elements of the mosquito which are: density; contact; longevity and competence of mosquitoes. Among other things, we can cite vector control without bednet use (home spraying, impregnated wall coverings, wire fencing, repellents, space sprays, genetic control) and the correct use of insecticide-treated mosquito bednets. Of all these methods, the possession and correct use of insecticide-treated mosquito bednets is the strategy that makes it possible to reduce three of the biological elements mentioned above \cite{Moiroux}. Therefore, we consider a control $ w (t) $ representing the effort made to own a mosquito net and to use it properly. Our optimal control malaria model consists of the following non-autonomous system of non-linear differential equations.
\begin{equation}
\left\lbrace \begin{array}{l}
S_h^{'}=\Lambda_h+\left( \gamma_h^{min}+w(t)\Delta_{\gamma_h}\right)  R_h-\left(\dfrac{m I_Q}{N_h}(a_{max}-w(t)\Delta_a)+\mu_h\right) S_h, \\ \\
I_h^{'}= \left(\dfrac{m I_Q}{N_h}(a_{max}-w(t)\Delta_a)\right) S_h-(\delta_h+\mu_h+\mu_d) I_h,\\ \\
R_h^{'}=\delta_h I_h-\left( \gamma_h^{min}+w(t)\Delta_{\gamma_h}+\mu_h\right) R_h ,\\ \\
S_q^{'}=\Lambda_v^{max}-w(t)\Delta_{\Lambda_v}-\left( \left( \dfrac{ c I_h}{N_h} +\dfrac{a \tilde{c} R_h}{N_h}\right)(a_{max}-w(t)\Delta_a) +\mu_v+w(t)\Delta_{\mu_v} \right) S_q, \\ \\
E^{{1}^{'}}_r=\left( \left( \dfrac{ c I_h}{N_h} +\dfrac{ \tilde{c} R_h}{N_h}\right)(a_{max}-w(t)\Delta_a) \right)S_q-(\chi+\mu_v)E^{1}_r, \\ \\
E^{{i}^{'}}_{q}=\chi E^{i}_{r}-(\beta+\mu_v+w(t)\Delta_{\mu_v}(t)) E^{i}_{q},\;\;\;\;\; 1 \leq i \leq 6, \\ \\
E^{{i}^{'}}_{r} =\beta E^{i-1}_{q}- (\chi+\mu_v) E^{i}_{r},\;\;\;\;\; 2 \leq i \leq 7,\\ \\
I_{r}^{'}=\beta I_q - (\chi+\mu_v)I_r,\\ \\
I_{q}^{'}=\chi (E_r^7+I_r)- (\beta+ \mu_v+w(t)\Delta_{\mu_v}(t))I_q,
\end{array} \right. \label{eqco}
\end{equation}

with initial conditions given at $t = 0$. Consider the following objective functional 
\begin{equation}
J(w)=\int_{0}^{T}\left[ A_1 I_h + A_2 \left( \sum^{6}_{i=1}E^{i}_{q}+ I_q\right)  +Bw^{2}(t)  \right] dt.  \label{foco}
\end{equation}

The term $A_1 I_h$ and $A_2 \left( \dsum_{i=1}^{6}E^{i}_{q}+ I_q\right)$ is the cost of infection while $Bw^{2}(t) $ is the cost of use of bednets. Our main goal is to find an optimal control function $w^{*}$ such that $J (w^{*}) = min\left\lbrace J(w) \mid w \in \Gamma(T)\right\rbrace $, with $\Gamma(T)$ the set of admissible controls, where $$\Gamma(T)=\left\lbrace \omega \mid \omega(.) \text{ is Lebesgue mesurable on}  \left[0,T \right], 0 \leq \omega(t) \leq 1  \text{ for  t in [0,T]} \right\rbrace . $$

The next step is to prove the existence of an optimal control for system \eqref{eqco} and then derive the optimality system.

\subsection{Existence of an optimal control}
\begin{thm}\label{optpb}
Consider the objective functional $J$ given by Equation \eqref{foco}, with $w \in \Gamma$ subject to the constraint state system \eqref{eqco}. There exists $w^{*} \in \Gamma(T)$ such that
$J (w^{*}) = min\left\lbrace J(w) \mid w \in \Gamma(T)\right\rbrace .$
\end{thm}
\begin{Proof}
Following similar results and the approach in \cite{abou}[Theorem 3.1., p.18], the proof is immediate.
\end{Proof}

\subsection{The optimality system}

To derive the necessary conditions that the three optimal controls and corresponding
states must satisfy, we use Pontryagin's maximum principle \cite{pont}. To this end, we define
the Hamiltonian function for the system, where $\lambda_i,\;\; i = 1,\ldots, 19$ are the adjoint variables or co-state variables
\begin{equation}
\begin{split}
\mathbb{H}&=A_1 I_h + A_2 \left( \sum^{6}_{i=1}E^{i}_{q}+ I_q\right)  +Bw^{2}(t) \\
&+\lambda_1 \left[ \Lambda_h+\left( \gamma_h^{min}+w(t)\Delta_{\gamma_h}\right)  R_h-\left(\dfrac{m I_Q}{N_h}(a_{max}-w(t)\Delta_a)+\mu_h\right) S_h \right\rbrace \\
&+ \lambda_2 \left[\left(\dfrac{m I_Q}{N_h}(a_{max}-w(t)\Delta_a)\right) S_h-(\delta_h+\mu_h+\mu_d) I_h \right] \\
&+ \lambda_3 \left[ \delta_h I_h-\left( \gamma_h^{min}+w(t)\Delta_{\gamma_h}+\mu_h\right) R_h \right] \\
&+ \lambda_4\left[\Lambda_v^{max}-w(t)\Delta_{\Lambda_v}-\left( \left( \dfrac{ c I_h}{N_h} +\dfrac{ \tilde{c} R_h}{N_h}\right)((a_{max}-w(t)\Delta_a) +\mu_v+w(t)\Delta_{\mu_v} \right) S_q \right] \\
&+ \lambda_5\left[ \left( \left( \dfrac{ c I_h}{N_h} +\dfrac{ \tilde{c} R_h}{N_h}\right)(a_{max}-w(t)\Delta_a) \right)S_q-(\chi+\mu_v)E^{1}_r \right] \\
&+\sum^{6}_{i=1} \lambda_{i+5} \left[ \chi E^{i}_{r}-(\beta+\mu_v+w(t)\Delta_{\mu_v}) E^{i}_{q}\right] \\ 
&+\sum^{7}_{i=2} \lambda_{i+10} \left[ \beta E^{i-1}_{q}- (\chi+\mu_v) E^{i}_{r} \right] \\ 
&+ \lambda_{18} \left[\beta I_q - (\chi+\mu_v)I_r \right] \\ 
&+ \lambda_{19} \left[\chi (E_r^7+I_r)- (\beta+ \mu_v+w(t)\Delta_{\mu_v})I_q \right]. \\ 
\end{split}
\end{equation}

The following result presents the adjoint system and control characterization.

\begin{thm}

\text{Given an optimal control } $w^* $, and corresponding state solutions $$S_h,I_h,R_h,S_q,E^1_r,E^1_q,E^2_q,E^3_q,E^4_q,E^5_q,E^6_q,E^2_r,E^3_r,E^4_r,E^5_r,E^6_r,E^7_r,I_r,I_q $$	 of the corresponding state system \eqref{eqc1}, there exists adjoint variables,	$\lambda_i,\;\; i = 1,\ldots, 19$, satisfying

\begin{equation}
\begin{small}
\left\lbrace \begin{array}{l}
\lambda'_1= \left(a_{max}-\Delta_aw(t) \right)\left[\dfrac{cI_h+\tilde{c}R_h}{N_h^{2}}S_q(\lambda_5-\lambda_4)+\dfrac{mI_q}{N_h}\left[\dfrac{S_h}{N_h}-1 \right](\lambda_2-\lambda_1)  \right]+\mu_h    \lambda_1,  \\ 
\lambda'_2=-A_1-\dfrac{mI_q}{N_h^2}\left(a_{max}-\Delta_aw(t) \right)S_h(\lambda_2-\lambda_1)+\dfrac{cN_h-cI_h-\tilde{c}R_h}{N_h^2}\left(a_{max}-\Delta_aw(t) \right)S_q(\lambda_4-\lambda_5)-\lambda_3 S_h -(\delta_h+\mu_h+\mu_d)\lambda_2, \\ 
\lambda'_3=(\gamma_h^{min}+w(t)\Delta_{\gamma_h}) (\lambda_3-\lambda_1)-\dfrac{mI_q}{N_h^2}\left(a_{max}-\Delta_aw(t) \right)S_h(\lambda_2-\lambda_1)+ \dfrac{\tilde{c}N_h-cI_h-\tilde{c}R_h}{N_h^2}\left(a_{max}-\Delta_aw(t) \right)S_q(\lambda_4-\lambda_5)+\mu_h\lambda_3,\\ 
\lambda'_4=(\mu_v+\tilde{\mu}w(t))\lambda_4-\dfrac{cI_h+\tilde{c}R_h}{N_h}\left(a_{max}-\Delta_aw(t) \right)(\lambda_5-\lambda_4), \\ 
\lambda'_5= (\chi + \mu_v) \lambda_5-\chi \lambda_6, \\ 
\lambda'_i=(\beta+\Delta{\mu_v}w(t))\lambda_i-\beta\lambda_{i+6}-A_2,\;\;\;for\;\;i=6,\ldots,11, \\ 
\lambda'_i=(\chi +\mu_v)\lambda_i -\chi \lambda_{i-5}\;\;\;for\;\;i=12,\ldots,16,  \\ 
\lambda'_{17}=(\chi+\mu_v)\lambda_{17}-\chi \lambda_{19},  \\ 
\lambda'_{18}=(\chi+\mu_v)\lambda_{18}-\chi \lambda_{19}, \\ 
\lambda'_{19}=(\beta +\mu_v+w(t)\Delta_{\mu_v})\lambda_{19}-\beta \lambda_{18}-\dfrac{m}{N_h}\left(a_{max}-\Delta_aw(t) \right)S_h(\lambda_2-\lambda_1)-A_2, \\ 
\end{array} \right. 
\end{small}
\end{equation}

with transversality conditions $\lambda_i(T) = 0,\;\;\;for\;\;i=1,\ldots,19 $ and the controls $w^{*} $ satisfy the optimality condition.

\begin{small}
\begin{equation}
w^{*} =max\left\lbrace 0,min \left(1,\dfrac{\Delta_{\gamma_h}R_h^{*}(\lambda_3-\lambda_1)+\Delta_a S_h^{*} \alpha_v^{*}(\lambda_2-\lambda_1)+\left(\Delta_a \alpha_h^{*}(\lambda_5-\lambda_4)+\Delta_{\mu_v}\lambda_4 \right)S_q^{*} +\lambda_{19}\Delta{\mu_v}I_q^{*} + \dsum_{i=1}^{6} \lambda_{i+5}E_{q}^{{i}^{*}}+\Delta_{\Lambda_{v}}\lambda_4}{2B} \right)  \right\rbrace , \label{optsol}
\end{equation}
\end{small}
where $\alpha_v^{*}=\dfrac{mI_q^{*}}{N_h^{*}}$  and $ \alpha_h^{*}=\dfrac{cI_h^{*}+\tilde{c}R_h^{*}}{N_h^{*}}.  $
\end{thm}

\begin{Proof}
The differential equations governing the adjoint variables are obtained by differentiation of the Hamiltonian function, evaluated at the optimal control. Then, the adjoint system can be written as

$$\lambda'_1(t)=-\dfrac{\partial \mathbb{H}}{\partial S_h},\;\;\;  \lambda'_2(t)=-\dfrac{\partial \mathbb{H}}{\partial I_h},\;\;\;  \lambda'_3(t)=-\dfrac{\partial \mathbb{H}}{\partial R_h},\;\;\;  \lambda'_4(t)=-\dfrac{\partial \mathbb{H}}{\partial S_q},\;\;\;  \lambda'_5(t)=-\dfrac{\partial \mathbb{H}}{\partial E_{r}^{1}},$$ 

$$\lambda'_i(t)=-\dfrac{\partial \mathbb{H}}{\partial E_{q}^{i-5}},\;\;for\;\;i=6,\ldots 11,\;\; \lambda'_i(t)=-\dfrac{\partial \mathbb{H}}{\partial E_{r}^{i-10}},\;\;for\;\;i=12,\ldots16,  $$
 
 $$\lambda'_{17}(t)=-\dfrac{\partial \mathbb{H}}{\partial E_{r}^{7}},\;\;\;  \lambda'_{18}(t)=-\dfrac{\partial \mathbb{H}}{\partial I_{r}},\;\;\;  \lambda'_{19}(t)=-\dfrac{\partial \mathbb{H}}{\partial I_{q}}, $$

with zero final time conditions (transversality) $\lambda_i(T)=0 $. The characterization of the optimal control given by \eqref{optsol} is obtained by solving the equations on the interior of the control set, where $0<w<1 $. That is, 
 $${\small \dfrac{\partial \mathbb{H}}{\partial \omega}=2B\omega^{*}-\left(\Delta_{\gamma_h}R_h^{*}(\lambda_3-\lambda_1)+\Delta_a S_h^{*} \alpha_v^{*}(\lambda_2-\lambda_1)+\left(\Delta_a \alpha_h^{*}(\lambda_5-\lambda_4)+\Delta_{\mu_v}\lambda_4 \right)S_q^{*} +\lambda_{19}\Delta{\mu_v}I_q^{*} + \dsum_{i=1}^{6} \lambda_{i+5}E_{q}^{{i}^{*}}+\Delta_{\Lambda_{v}}\lambda_4 \right)}, $$ with $\dfrac{\partial \mathbb{H}}{\partial \omega}=0$,
where $\alpha_v^{*}=\dfrac{mI_q^{*}}{N_h^{*}}$  and $ \alpha_h^{*}=\dfrac{cI_h^{*}+\tilde{c}R_h^{*}}{N_h^{*}}.$  Hence, we obtain $$\omega^{*}=\dfrac{\Delta_{\gamma_h}R_h^{*}(\lambda_3-\lambda_1)+\Delta_a S_h^{*} \alpha_v^{*}(\lambda_2-\lambda_1)+\left(\Delta_a \alpha_h^{*}(\lambda_5-\lambda_4)+\Delta_{\mu_v}\lambda_4 \right)S_q^{*} +\lambda_{19}\Delta{\mu_v}I_q^{*} + \dsum_{i=1}^{6} \lambda_{i+5}E_{q}^{{i}^{*}}+\Delta_{\Lambda_{v}}\lambda_4}{2B}.$$
\end{Proof}

\section{Numerical simulations: the biological admissibility and approximate controls}\label{sec4}

We numerically solve the optimal transmission parameter control for the malaria model. The optimal control is obtained by solving the optimality system, consisting of $19$ non-linear ordinary differential equations from the state and adjoint equations. An iterative scheme is used for solving the optimality system \cite{len}. In simulation, we consider the initial number of individuals at $t=0$: $S_h(0)= 100000, I_h(0)=100, R_h(0)=1000,Sq(0)=100000$, $E^{1}_r(0)=10$, $E^{2}_r(0)=9,E^{3}_r(0)=8$, $E^{4}_r(0)=7,E^{5}_r(0)=6,E^{6}_r(0)=5,E^{7}_r(0)=4$, $E^{1}_q(0)=3,E^{2}_q(0)=3,E^{3}_q(0)=3,E^{4}_q(0)=3$, $E^{5}_q(0)=3$, $E^{6}_q(0)=2,I_r(0)=35,I_q(0)=800$.

For the cost weight in the objective functional $J$, we take $B=\$4.5\;\;USD $ (for three years) which represents what the state of Cameroon spends on the purchase of an insecticide-treated mosquito net \cite{pul} for two individuals. It is possible to compare to the $B=\$3.95\;\; USD$ of \cite{ndo} for the average cost for a household (with about 5.5 individuals) per month (for the first two largest cities of Cameroon in terms of population - Douala and Yaounde). The main practical problem is the difficulty to provide bednets to everybody in the household as well as individuals complain of feeling excessive heat when sleeping under a bednet\cite{ndo}, the latter being a potential reason  why some individuals use other vector control measures.  We could consider the cost per $household^2$ for $T=3 years$, and finally discuss between the effects of the optimal controls associated to $B_1=4.5*\frac{5.5}{2}$ and $B_2=3.95*36$. Clearly, from the economic stand point, the other vector control strategies are more expensive than the bednet control. Thus, we focus on the bednet control strategies. The simulations are carried out with $T$ = three years, the duration of one LLIN (Long-lasting insecticide-treated bednet) \cite{pul} efficacy. We define the uniform control $u_{unif}(t)=k$ and the multi-intervals ("stage") one $u_{stages}(t)=\left\lbrace \begin{array}{ccc}
u_1 & • & t\in\left[0;365.25\textbf{ days } \right],  \\ 
u_2 & • & t\in\left]365.25;730.5\textbf{ months } \right], \\ 
u_3 & • & t\in\left]730.5;1080.75\textbf{ months } \right],
\end{array} \right. $ over three years with $k,u_1,u_2,u_3\in\mathbb{R_+}$ and $\frac{1}{1080.75}\int^{1080.75}_0 u_{stages}(t)dt =k$ as the mean value. $u_{optimal}$ is an optimal control for our optimal problem in Theorem \ref{optpb}. $u_{forced}$ is an administrative control of distribution of the bednets over three years; it is either $u_{unif}$ or $u_{stages}$. We also define the following in percentage:
\begin{itemize}
\item[1.] $t^{u_{optimal}}_s(u_{forced})=\frac{100\times\textbf{Total of susceptible humans on [ 0;T] under }u_{forced} }{\textbf{ Total of susceptible humans on [ 0;T] under }u_{optimal}}$;
\item[2.] $t^{u_{optimal}}_{I_h}(u_{forced})=\frac{100\times\textbf{Total of infectious humans on [ 0;T] under }u_{optimal}}{\textbf{ Total of susceptible humans on [ 0;T] under }u_{forced}}$;
\item[3.] $t^{u_{optimal}}_{R_h}(u_{forced})=\frac{100\times\textbf{Total of recovered humans on [ 0;T] under }u_{optimal} }{\textbf{Total of recovered humans on [ 0;T] under }u_{forced}}$.
\end{itemize}
\begin{definition}\label{def} Let $\Gamma(T)$ be the set of admissible controls relative to a dynamical system $D_{(u(.))} $, $u(.)\in\Gamma(T)$.
An optimal control, mathematically admissible, is \textbf{biologically admissible} to $u_{forced}$ if $t^{u_{optimal}}_s(u_{forced})\leq 100$, $t^{u_{optimal}}_{I_h}(u_{forced})\leq 100$ and $t^{u_{optimal}}_{R_h}(u_{forced})\leq 100$.

An optimal control, mathematically admissible, is \textbf{biologically admissible} if $$t^{u_{optimal}}_s(u_{forced})\leq 100$$ $$t^{u_{optimal}}_{I_h}(u_{forced})\leq 100$$ and $$t^{u_{optimal}}_{R_h}(u_{forced})\leq 100$$ for all mathematically admissible control $u_{forced}$.
\end{definition}

It is easy (even numerically) to study the biological admissibility to an (mathematically) admissible control $u_{forced}$. But (for all $u_{forced}$) the biological admissibility is a challenge related to the choice of the objective function.

\begin{figure}[H]
\begin{center}
\includegraphics[scale=0.4]{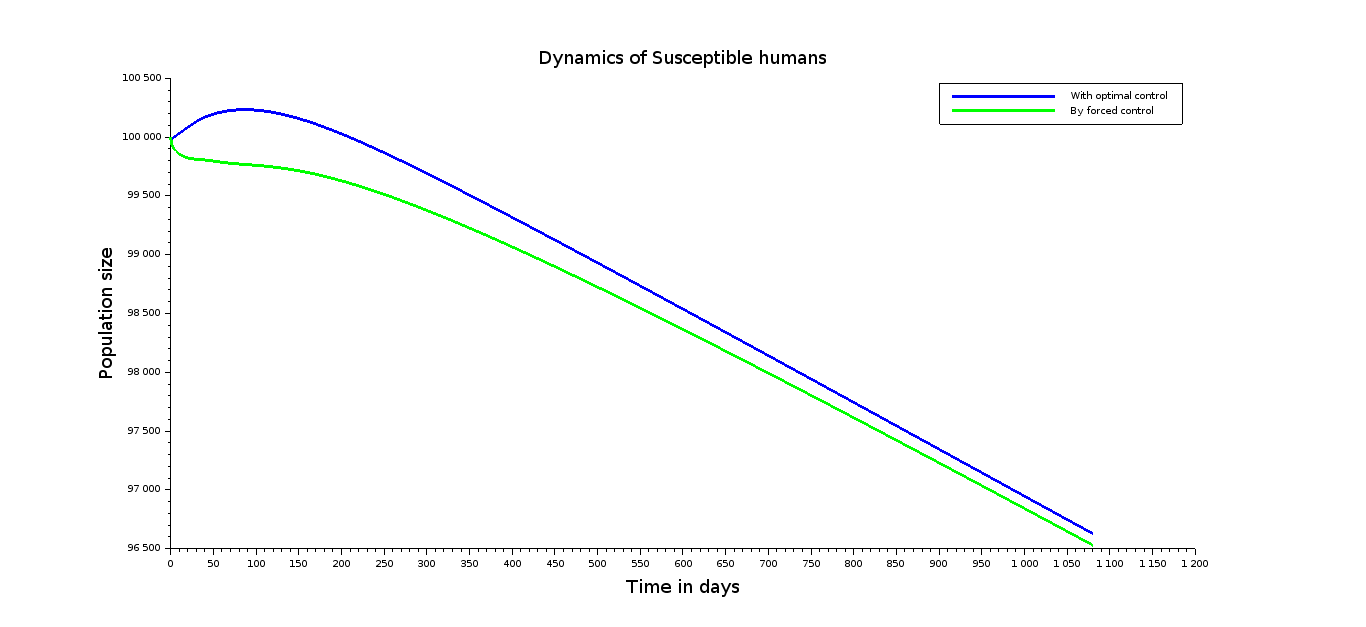} 
\caption{The number of susceptible humans $S_h$: optimal versus uniform controls}
\label{unif1}
\end{center} 
\end{figure}

\begin{figure}[H]
\begin{center}
\includegraphics[scale=0.4]{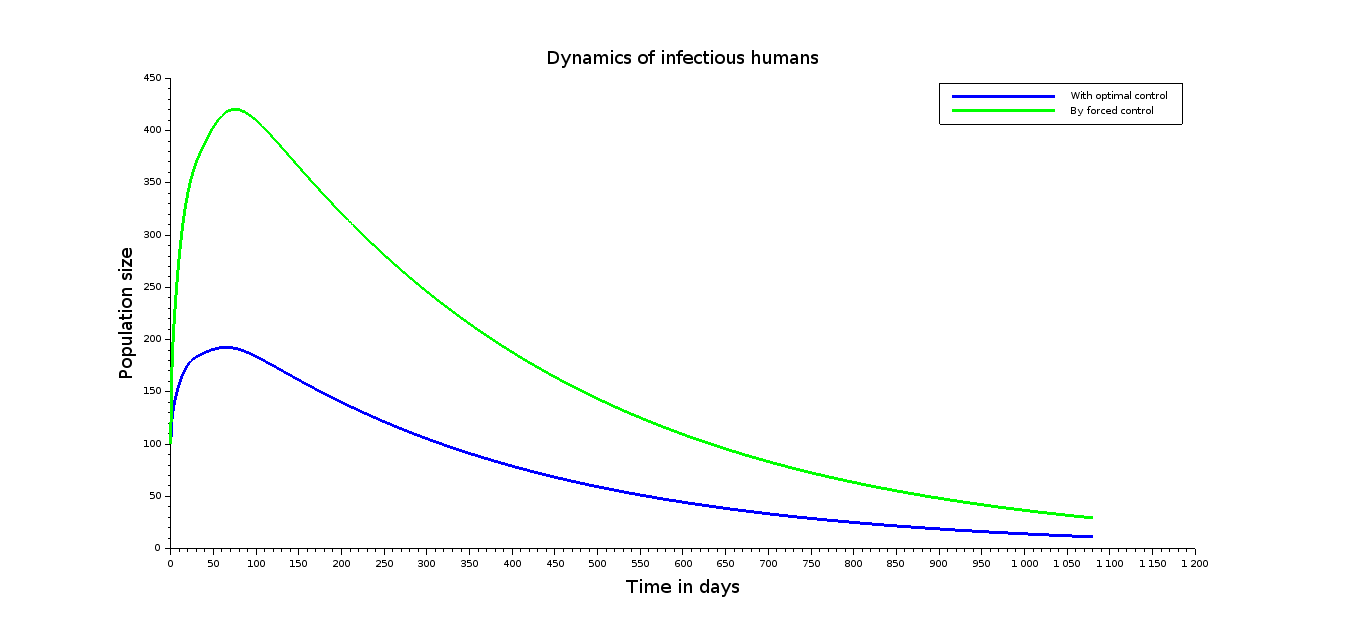} 
\caption{The number of infectious humans $I_h$: optimal versus uniform controls}
\label{unif2}
\end{center} 
\end{figure}

\begin{figure}[H]
\begin{center}
\includegraphics[scale=0.3]{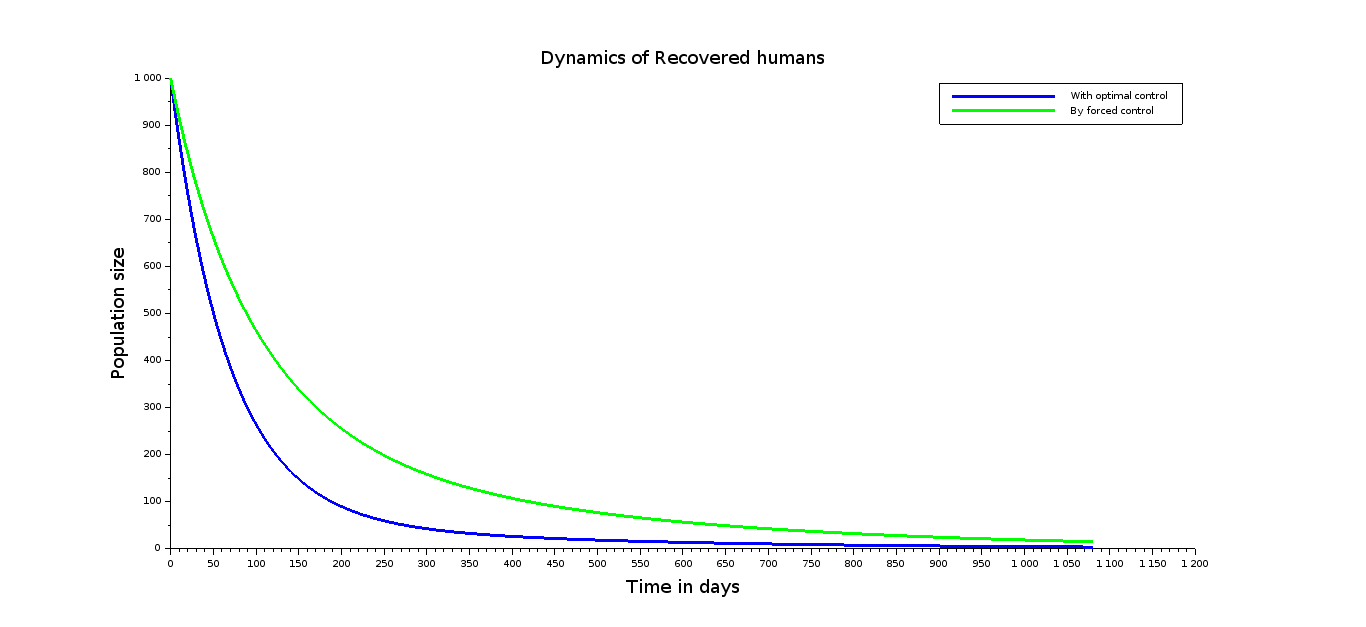} 
\caption{The number of recovered individuals $R_h$: optimal versus uniform controls}
\label{unif3}
\end{center} 
\end{figure}

\begin{figure}[H]
\begin{center}
\includegraphics[scale=0.4]{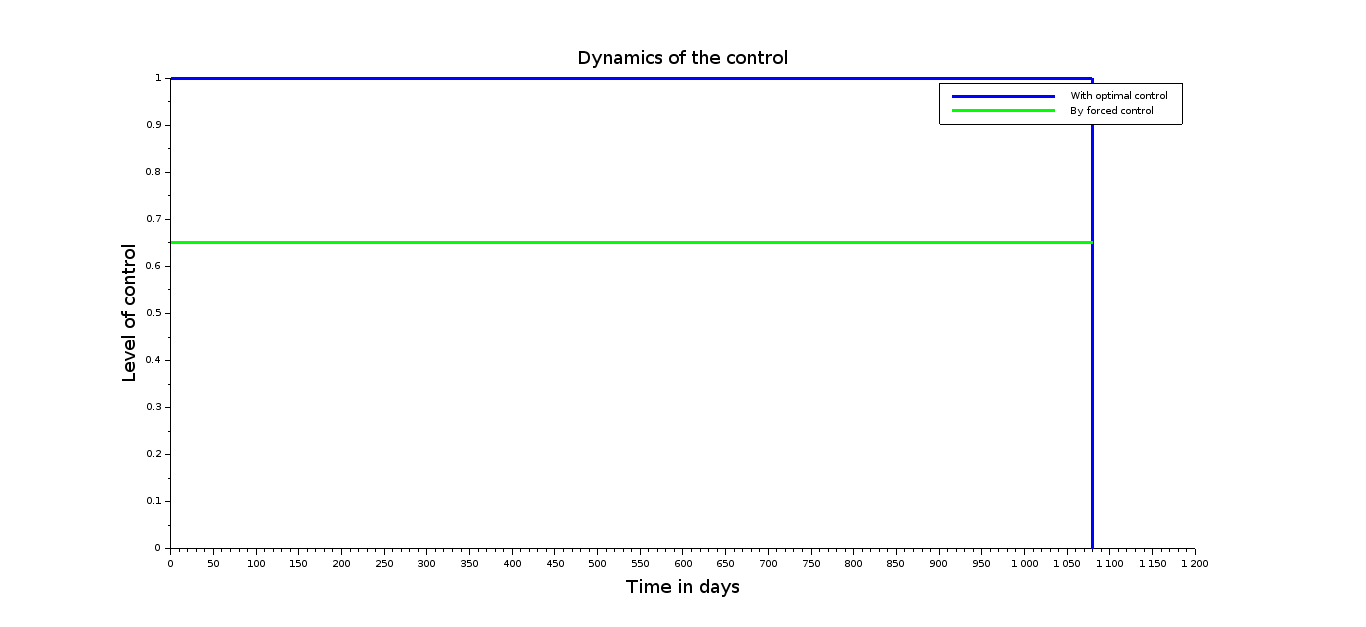} 
\caption{The optimal control $u_{optimal}$ compared to the uniform control $u_{unif}$: optimal versus uniform controls}
\label{unif4}
\end{center} 
\end{figure}

\begin{figure}[H]
\begin{center}
\includegraphics[scale=0.3]{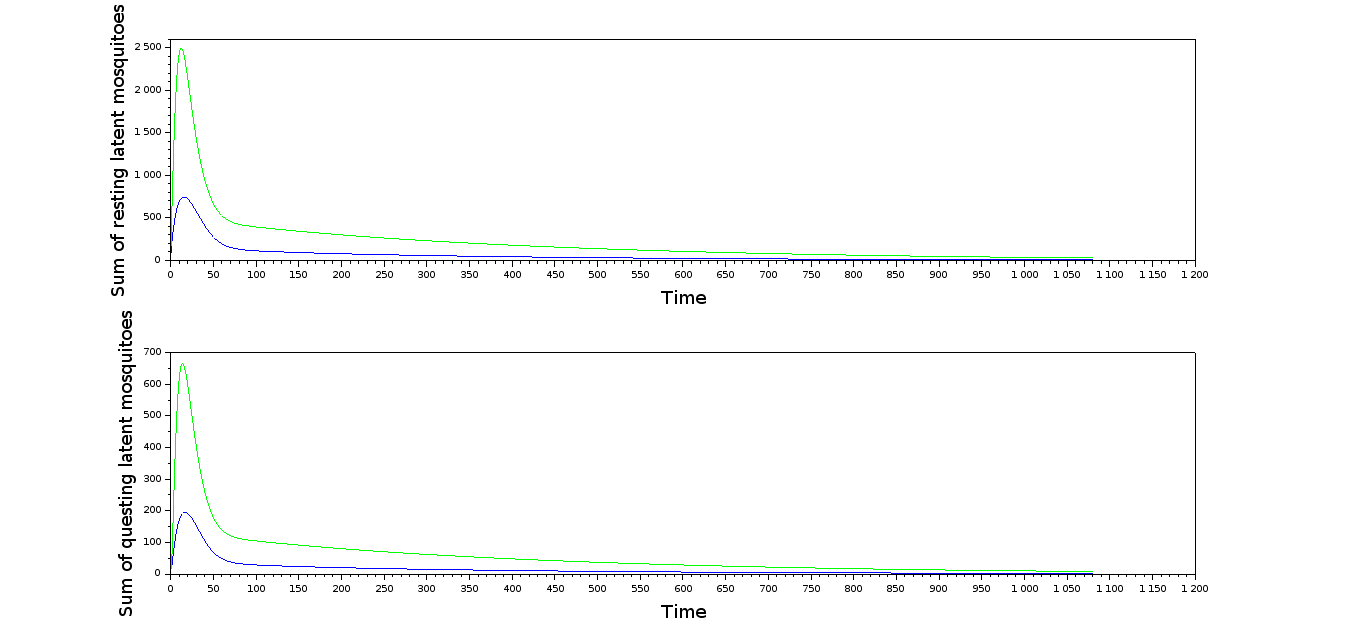} 
\caption{The total number of latent questing $E^{.}_q$ and latent resting $E^{.}_r$ mosquitoes: optimal versus uniform controls}
\label{unif5}
\end{center} 
\end{figure}

\begin{figure}[H]
\begin{center}
\includegraphics[scale=0.3]{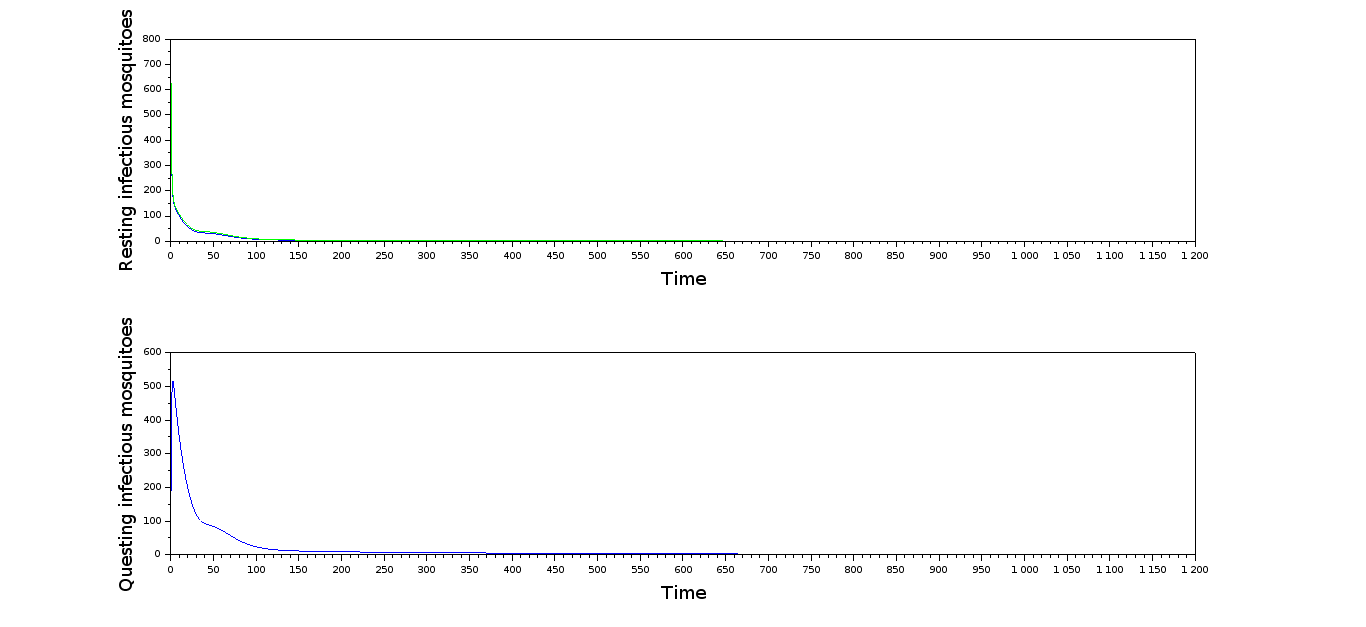} 
\caption{The number of questing $I_q$ and resting $I_r$ infectious mosquitoes: optimal versus uniform controls}
\label{unif6}
\end{center} 
\end{figure}

\begin{figure}[H]
\begin{center}
\includegraphics[scale=0.3]{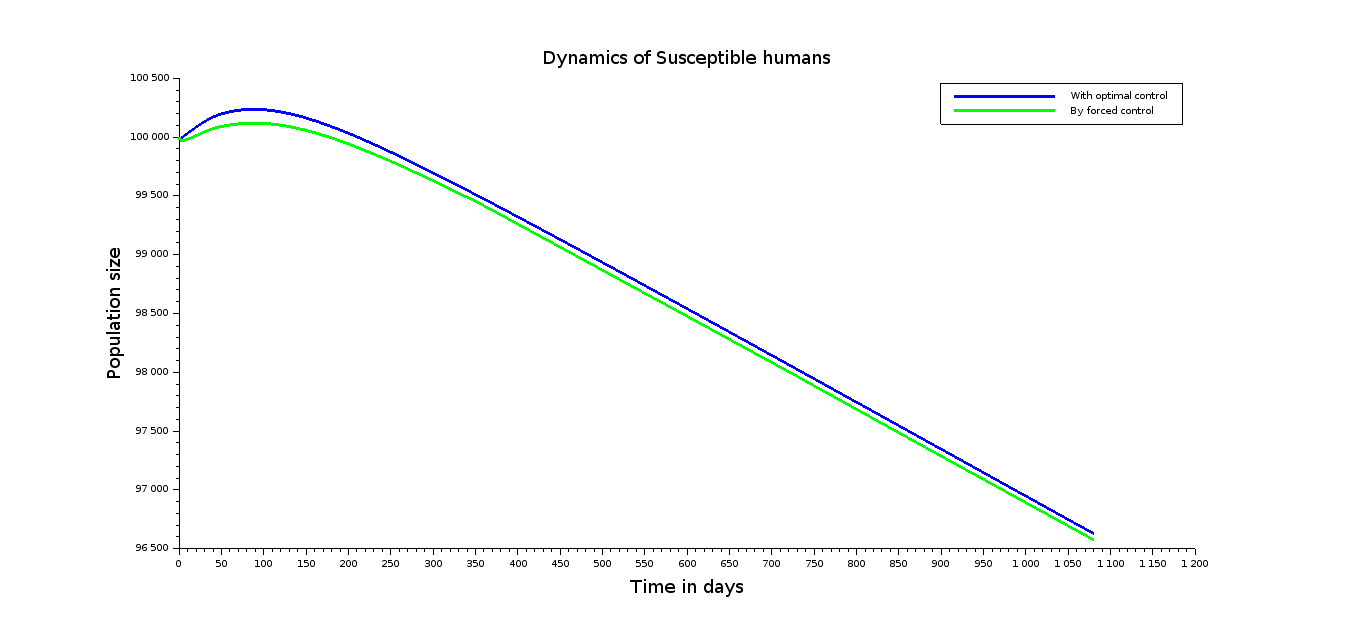} 
\caption{The number of susceptible humans $S_h$: optimal versus "stage" controls}
\label{stage1}
\end{center} 
\end{figure}

\begin{figure}[H]
\begin{center}
\includegraphics[scale=0.4]{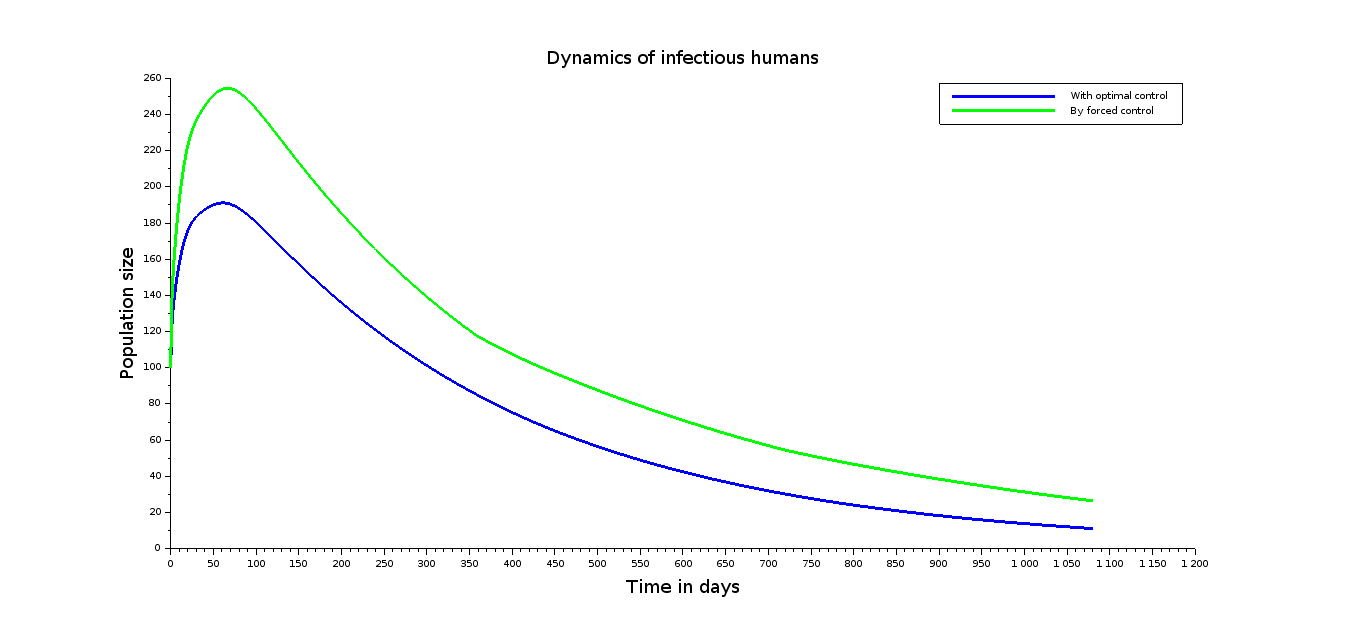} 
\caption{The number of infectious humans $I_h$: optimal versus "stage" controls}
\label{stage2}
\end{center} 
\end{figure}

\begin{figure}[H]
\begin{center}
\includegraphics[scale=0.4]{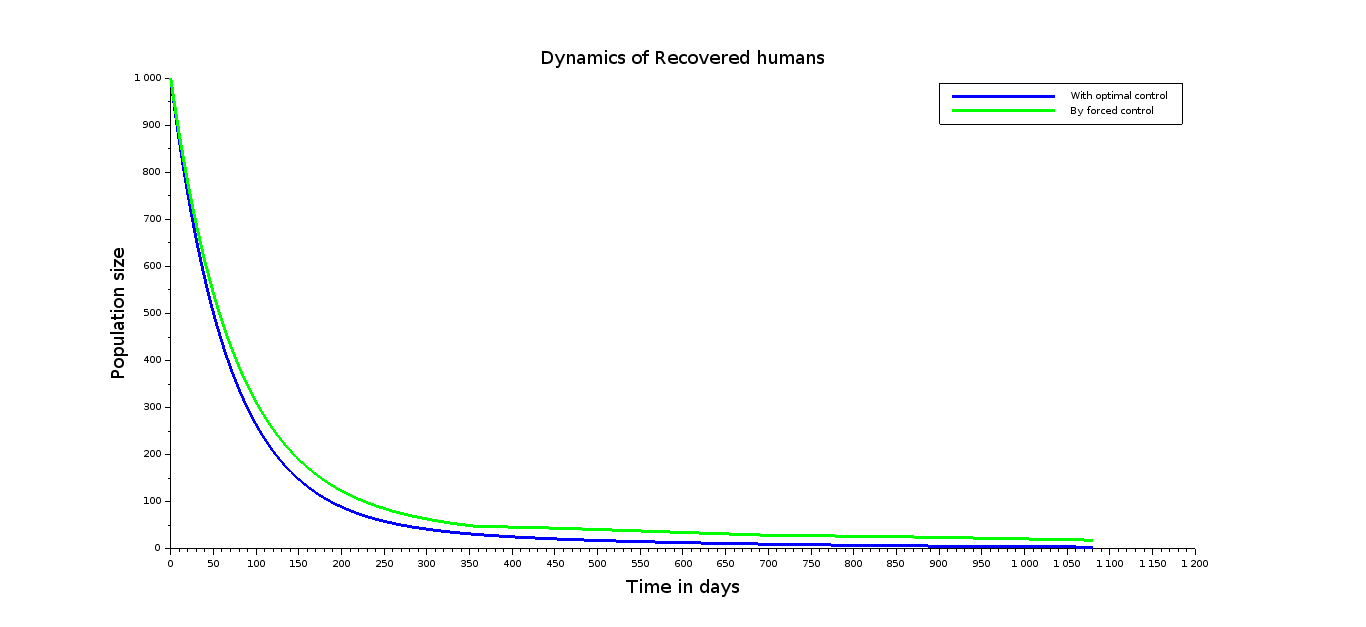} 
\caption{The number of recovered individuals $R_h$: optimal versus "stage" controls}
\label{stage3}
\end{center} 
\end{figure}

\begin{figure}[H]
\begin{center}
\includegraphics[scale=0.3]{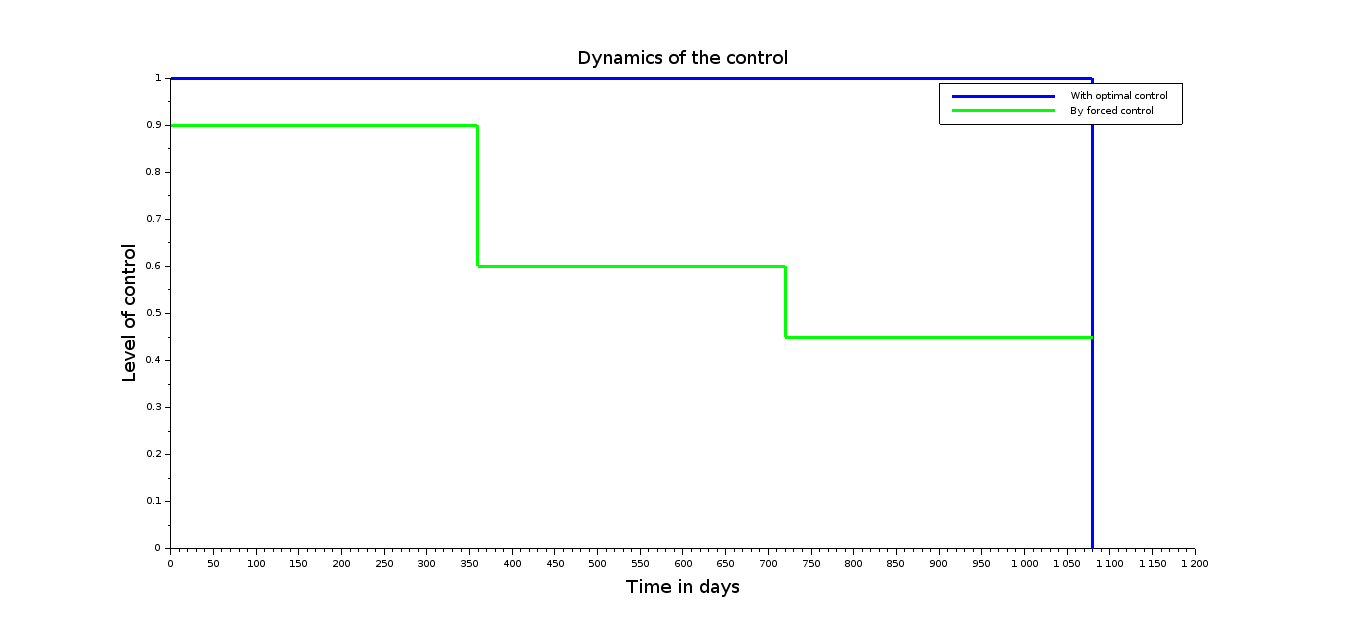} 
\caption{The optimal control $u_{optimal}$ compared to the "stage" control $u_{stage}$}
\label{stage4}
\end{center} 
\end{figure}

\begin{figure}[H]
\begin{center}
\includegraphics[scale=0.25]{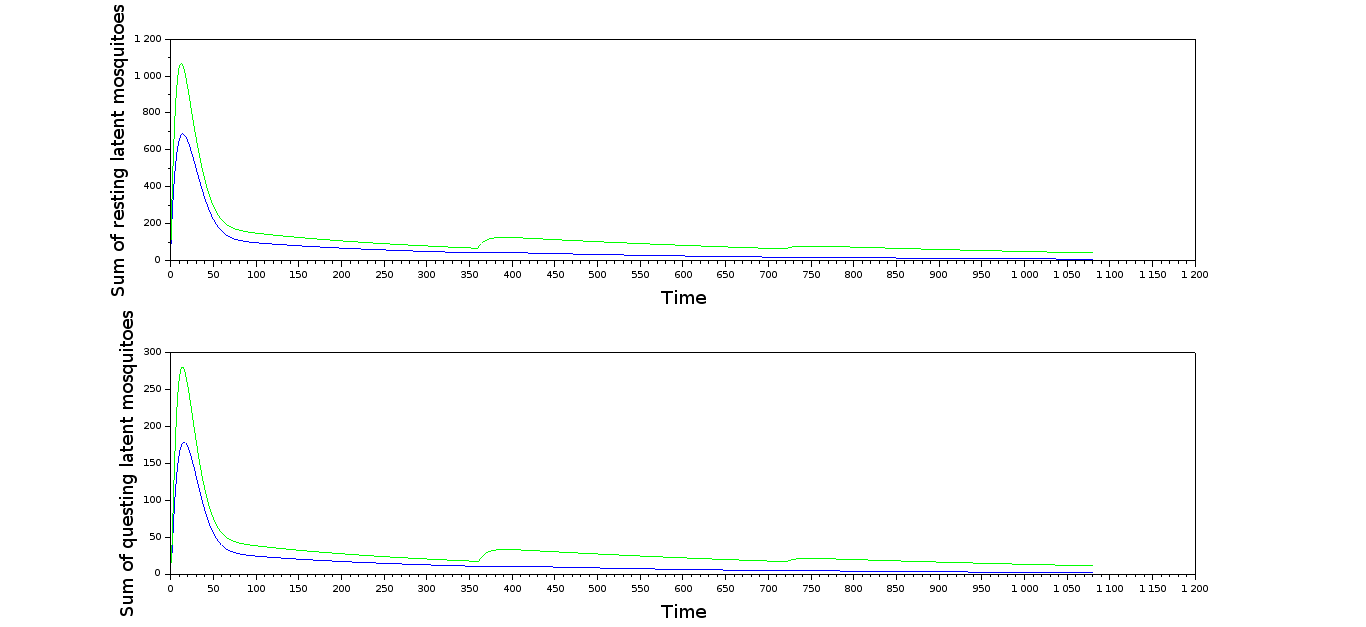} 
\caption{The total number of latent questing $E^{.}_q$ and latent resting $E^{.}_r$ mosquitoes: optimal versus "stage" controls}
\label{stage5}
\end{center} 
\end{figure}

\begin{figure}[H]
\begin{center}
\includegraphics[scale=0.3]{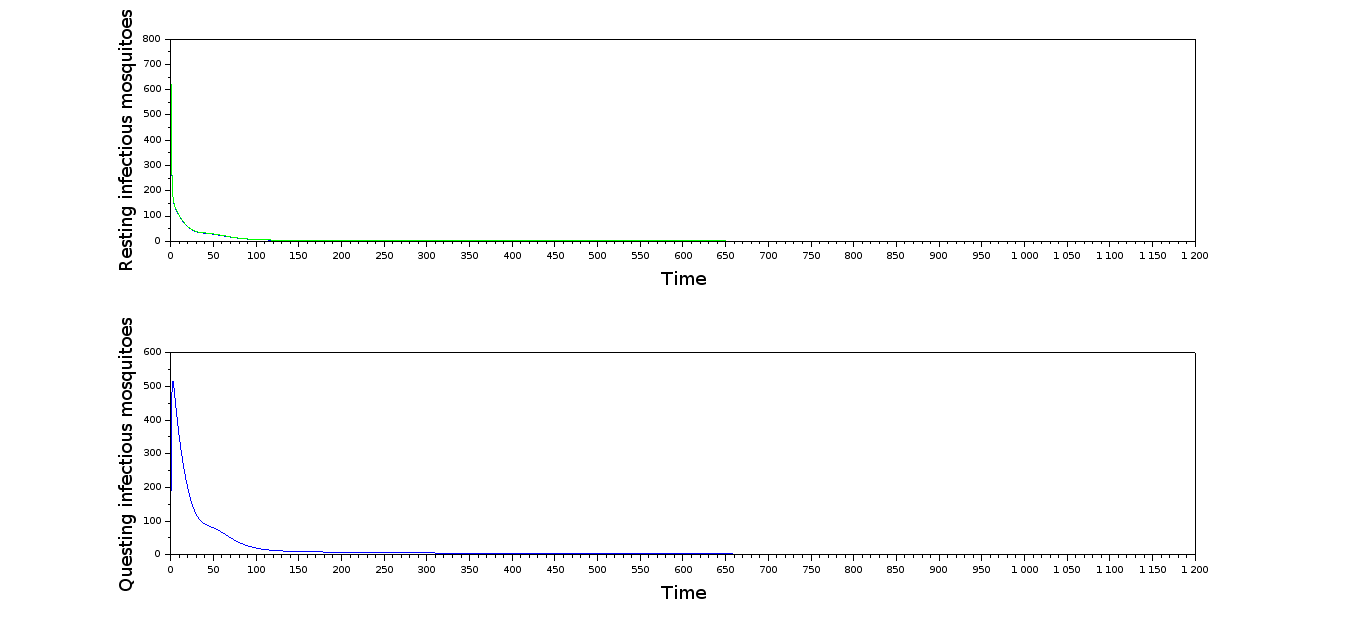} 
\caption{The number of questing $I_q$ and resting $I_r$ infectious mosquitoes: optimal versus "stage" controls}
\label{stage6}
\end{center} 
\end{figure}

Numerically, for $T=1080.75$ days, $u_{unif}(t)=0.65$ and $u_{stages}(t)=\left\lbrace \begin{array}{lll}
u_1=0.9 &  & t\in\left[0;365.25\textbf{ days } \right],  \\ 
u_2=0.6 &  & t\in\left]365.25;730.5\textbf{ days } \right], \\ 
u_3=0.45 &  & t\in\left]730.5;1080.75\textbf{ days } \right]. 
\end{array} \right. $ For all our numerical simulations, graphs related to optimal control are in black solid lines, while those linked to the "forced" control are in solid green lines. The effects of the "uniform" control are graphically represented in Figures \ref{unif1}-\ref{unif6} while the "stage" control effects are shown in Figures \ref{stage1}- \ref{stage6}.

Practically, the common strategies $u_{forced}$ in malaria affected countries are decreasing functions of time (as trends), due to the difficulty to maintain a constant or high ($>90\%$) level of possession and use of bednets throughout the 3- year campaign of LLINs distribution.
\begin{table}[H]\label{tab2}
\caption{Results in percentage}
\begin{center}
\begin{tabular}{c|c|c|c}
\hline 
 & $t^{u_{optimal}}_s(u_{forced})$ & $t^{u_{optimal}}_{I_h}(u_{forced})$ & $t^{u_{optimal}}_{R_h}(u_{forced})$ \\ 
\hline 
$u_{unif}$ & 99.768734 & 43.529843 & 52.163645 \\ 
\hline 
$u_{stages}$ & 99.929462 & 68.500328 & 77.009369 \\ 
\hline 
\end{tabular} 
\end{center}
\end{table}

Table \ref{tab2} suggests that, even if the "stage" and uniform controls have the same mean, it is better to use a "stage" control with emphasis on the first few months of the 3 years.
Clearly, the effort should be done to cover the gap (between the results of the system state following $u_{forced}$ and $u_{optimal}$) in Table \ref{tab2}. This approach, based on the reality of the malaria programs in each country, could support health policies and  decision-makers in order to obtain an accurate threshold in the percentage $\epsilon_{u_{forced}}$ of the "administrative/public planners controls" $u_{forced}$ ($u_{stage}$ or $u_{unif}$) applications compared to optimal effects, such that $100-t^{u_{optimal}}_s(u_{forced})\leq \epsilon_{u_{forced}}$, $100-t^{u_{optimal}}_{I_h}(u_{forced})\leq \epsilon_{u_{forced}}$ and $100-t^{u_{optimal}}_{R_h}\leq \epsilon_{u_{forced}}$. This allows us to introduce the definitions of the $\epsilon_{u_{forced}}$-approximate weak or strong "sub-optimal" controls.
\begin{definition}\label{def2} (Approximate controlability) Let $\Gamma(T)$ be the set of admissible controls relative to a dynamical system $D_{(u(.))} $, $u(.)\in\Gamma(T)$, for $T>0$,. For $V^{u_{optimal}}_{u_{forced}}=\left(t^{u_{optimal}}_s(u_{forced}), t^{u_{optimal}}_{I_h}(u_{forced}),t^{u_{optimal}}_{R_h}(u_{forced})\right) $, let define $$Norm_{strong}(V^{u_{optimal}}_{u_{forced}}):=max\left\lbrace 100-t^{u_{optimal}}_s(u_{forced}),100-t^{u_{optimal}}_{I_h}(u_{forced}), 100-t^{u_{optimal}}_{R_h}(u_{forced})\right\rbrace, $$ also written as $$Norm_{strong}(V^{u_{optimal}}_{u_{forced}}):=100-min\left\lbrace t^{u_{optimal}}_s(u_{forced}),t^{u_{optimal}}_{I_h}(u_{forced}), t^{u_{optimal}}_{R_h}(u_{forced})\right\rbrace, $$ and $$Norm_{weak}(V^{u_{optimal}}_{u_{forced}}):= \frac{1}{3}\left( 100-t^{u_{optimal}}_s(u_{forced})\right) +\left( 100-t^{u_{optimal}}_{I_h}(u_{forced})\right) +\left( 100-t^{u_{optimal}}_{R_h}\right).$$ That is, $$ Norm_{weak}(V^{u_{optimal}}_{u_{forced}}):= 100-\frac{1}{3}\left\lbrace t^{u_{optimal}}_s(u_{forced})+t^{u_{optimal}}_{I_h}(u_{forced})+t^{u_{optimal}}_{R_h}\right\rbrace. $$

A biologically admissible control $u_{forced}$ is $\epsilon_{u_{forced}}$-approximate weak "sub-optimal" if $$Norm_{weak}({V^{optimal}}(u_{forced}))\leq \epsilon_{u_{forced}}.$$

A biologically admissible control $u_{forced}$ is $\epsilon_{u_{forced}}$-approximate strong "sub-optimal" if $$Norm_{strong}(V^{u_{optimal}}_{u_{forced}})\leq \epsilon_{u_{forced}}.$$
\end{definition}

\begin{rmqs}
These definitions in \ref{def2}  improve on the efficiency index \cite{abou}. It is possible to consider the reduction of noise $N_{mosq}$ (similar to  $Norm_{.}$ for mosquitoes) produced by mosquitoes as the percentage of mosquitoes with optimal control compared to the states with forced control: then the new index would be $ Norm^{\alpha,\beta}_{.}:=\alpha Norm_{.}+\beta N_{mosq}$ such that $\alpha+\beta=1$. The coefficients $ \alpha$ and $\beta$ traduced respectively the importance of the humans' group and mosquitoes' group. In this paper, we focus on the optimal impact on humans and consider $\alpha = 1$.
\end{rmqs}
straithforward computations lead to this proposition.
\begin{prop}
There is an equivalence between $Norm_{weak}$ and $Norm_{weak}$: $$ Norm_{weak} \leq Norm_{strong} \leq 3.Norm_{weak}$$
\end{prop}

In our case in Table \ref{tab2}, $$Norm_{weak}(V^{u_{optimal}}_{u_{unif}})= 34.845926,$$ $$Norm_{weak}(V^{u_{optimal}}_{u_{stage}})= 18.186947,$$ $$Norm_{strong}(V^{u_{optimal}}_{u_{unif}})= 56.470157,$$ and $$Norm_{strong}(V^{u_{optimal}}_{u_{stage}})= 31.499672.$$ We see that for $\epsilon_{u_{forced}}=35\%$, the $V^{u_{optimal}}_{u_{unif}}$ is 35$\%$-approximate weak "sub-optimal" like $V^{u_{optimal}}_{u_{stage}}$. But only $V^{u_{optimal}}_{u_{stage}}$ is 35$\%$-approximate strong "sub-optimal" and not $V^{u_{optimal}}_{u_{unif}}$. Another interesting point is the fact that $Norm_{weak}(V^{u_{optimal}}_{u_{stage}})= 18.186947$, and this comes from the fact that the "weak" deviation from the optimal strategy is only about $18.186947\%$ in total (with the collective effort/contribution of $S_h, I_h$ and $R_h$ to reach the optimal strategy). By the way, $Norm_{strong}(V^{u_{optimal}}_{u_{stage}})= 31.499672$ and this corroborates the fact that the "weak" deviation from the optimal strategy is only about $31.499672\%$ following the \textbf{individual} efforts/contribution of $S_h, I_h$ and $R_h$ to reach the optimal objective.

\section{Conclusion}\label{sec5}
We formulated and rigorously analyzed a vector multi-stage malaria model with the use of mosquitoes treated bednets as preventive measure. The proposed model is biologically meaningful and mathematical well-posed. We investigated the local and global stability of equilibria. The analytical results reveal the possibility of bistability when $ \mathcal{R}_0 <\zeta <1$ (see subsection \ref{bf} with additional mortality $\delta_h$ less than $10^{-5}$ see discussion in \cite{Agusto}). That is, the model could exhibit the phenomenon of backward bifurcation, an epidemiological situation where although necessary, having the basic reproduction number less than unity is not sufficient to mitigate the malaria transmission dynamics \cite{jmt}. Thus, a low level of additional (disease-induced) mortality could lead to the existence of an endemic equilibrium even if the basic reproduction rate is less than one. Next, an optimal control strategy is investigated with the correct usage of LLINs (during three years compared to a "forced" control) as the control parameter. Results from this study could help inform health policy and decision-makers on the potential optimum strategies to mitigate malaria transmission dynamics in affected communities by designing reachable malaria program implementation objectives "close" to the optimal strategies $\epsilon_{u_{.}}\%$ by the "weak" collective contribution or the "strong" individual effort to achieve the optimal objective. The notions of $\epsilon_{u_{forced}}$-approximate strong/weak "sub-optimal" control are more practical than the theoretical optimal control which remains a daunting task to health officials. The upper bound $\epsilon_{u_{forced}}$ of the gap, from the "sub-optimal" results to the optimal ones, is of great interest practically since it delineates the acceptable error one could essentially make if we apply $u_{forced}$ instead of $u_{optimal}$.
 
\bibliographystyle{plain}
\bibliography{BiblioBiomath}
\end{document}